\documentclass[letterpaper,showpacs,onecolumn,eqsecnum,superscriptaddress,floatfix,nofootinbib]{revtex4}

\usepackage{amssymb}
\usepackage{amsmath}
\usepackage{latexsym}
\usepackage{epsfig}
\usepackage{graphicx}
\usepackage{color}
\usepackage[linktocpage]{hyperref}
\usepackage{epstopdf}
\usepackage[english]{babel}

\begin{document}

%%%%%%%%%%%%%%%%%
%%%   TITLE   %%%
%%%%%%%%%%%%%%%%%
%
\title{Obtaining Gravitational Waves from Inspiral Binary Systems using LIGO data}

%%%%%%%%%%%%%%%%%
%%%  AUTHORS  %%%
%%%%%%%%%%%%%%%%%
%
\author{Javier M. Antelis} 
\email[]{mauricio.antelis@itesm.mx}
\affiliation{Tecnol\'ogico de Monterrey, Campus Guadalajara, \\
Av. Gral. Ram\'on Corona 2514, Zapopan, Jal., 45201, M\'exico}

\author{Claudia Moreno} 
\email[]{claudia.moreno@cucei.udg.mx} 
\affiliation{Departamento de F\'isica,
Centro Universitario de Ciencias Exactas e Ingenier\'ias, Universidad de Guadalajara,\\
Av. Revoluci\'on 1500, Colonia Ol\'impica C.P. 44430, Guadalajara, Jalisco, M\'exico}
\affiliation{Instituto de F\'isica y Matem\'aticas,
Universidad Michoacana de San Nicol\'as de Hidalgo,\\
Edificio C-3, Ciudad Universitaria, 58040 Morelia, Michoac\'an, M\'exico.}

%%%%%%%%%%%%%%%%
%%%   DATE   %%%
%%%%%%%%%%%%%%%%
%
%\date{\today}

%%%%%%%%%%%%%%%%%%%%
%%%   ABSTRACT   %%%
%%%%%%%%%%%%%%%%%%%%
%
\bigskip
\begin{abstract}
%%This work presents a brief introduction to gravitational waves and describes the scientific efforts devoted to detect them using the laser interferometer gravitational wave observatory (LIGO). \CM {poner un objetivo especifico mas}. Results of detection of gravitational waves from inspiral compact binaries injected in LIGO's fifth and sixth science runs are presented and discussed.
%%
%We analyzed GW generated by inspiral compact binaries, a system composed of a pair of neutron stars in their late stage of evolution. Also, we perform the detection of this type of GW injected in the freely available LIGO data. We present a comprehensive data analysis methodology based on the Matched Filter, which is the optimal detection technique of GW emitted by these astrophysical sources. Finally, the results of the detection experiments are presented and discussed.
%
%\textit{Abstract}: 
The discovery of the events GW150926 and GW151226 has experimentally confirmed the existence of gravitational waves (GW) and has demonstrated the existence of binary stellar-mass black hole systems. This finding marks the beginning of a new era that will reveal unexpected features of our universe. This work presents a basic insight to the fundamental theory of GW emitted by inspiral binary systems and describes the scientific and technological efforts developed to measure this waves using the interferometer-based detector called LIGO. Subsequently, the work proposes a comprehensive data analysis methodology based on the matched filter algorithm which aims to detect GW signals emitted by inspiral binary systems of astrophysical sources. The method is validated with freely available LIGO data which contain injected GW signals. Results of experiments performed to assess detection carried out show that the method was able to recover the 85\% of the injected GW.
%\\
%\textit{Keywords}: Gravitational Waves; LIGO; Detection Algorithm; Data Analysis.
\end{abstract}

%\begin{keyword}
%Gravitational Waves, LIGO, Inspiral Binary, Detection Algorithm, Data Analysis
%\end{keyword}

%%%%%%%%%%%%%%%%
%%%   PACS   %%%
%%%%%%%%%%%%%%%%
%
\pacs{
04.30.-w, % gravitational waves
07.05.Kf,  % data analysis    
95.30.Sf,  % relativity and gravitation
11.15.Bt % perturbation theory 
}
% 04.25.-g

%

%%%%%%%%%%%%%%%%%%%%%%
%%%   MAKE TITLE   %%%
%%%%%%%%%%%%%%%%%%%%%%
%
\maketitle

%%%%%%%%%%%%%%%%%%%%%%%%
%%%     SECTIONS     %%%
%%%%%%%%%%%%%%%%%%%%%%%%
%
%\input{01_Introduction}
%%%%%%%%%%%%%%%%%%%%%%%%%%%%%%%%%%%%%%%
\section{Introduction}
\label{sec:introduction}
The recent announcements of the first observations of Gravitational Waves (GW) have caused great excitement and enthusiasm in the scientific community since this discovery opens a new window to explore our universe \cite{Holst2016}.
GW are ripples of the spacetime geometry created by massive astrophysical objects which were predicted by Albert Einstein one hundred years ago. 
%
% GW are one of the four fundamental forces in the universe \cite{FForces}.
%
In 1915, Einstein presented the theory of General Relativity, a set of nonlinear partial differential equations which basically says that mass and energy produce a curvature on the four-dimensional spacetime geometry and matter moves in response to that curvature \cite{Einstein15a}. % mass and energy tells spacetime how to bend and curved spacetime tells mass how to move
One year later, Einstein linearized and simplified the equations which revealed the existence of GW \cite{Einstein15b}.
These oscillations of the spacetime metric were initially regarded a mere theoretical aspect until the observations of the binary pulsar PSR 1913+16 performed by astronomers Russell Hulse and Joseph Taylor during 1970s and 1980s showed indirect evidence of its existence \cite{rHjT75,jTjW04}.
They noticed that the orbital period of this astrophysical object is gradually decreasing and that its energy loss is in agreement with the General Relativity prediction of GW emission~\cite{jTjW82,tDjT92}. %They noticed that the orbital period of this astrophysical object, which is composed of two neutron stars spiralling inwards, is gradually decreasing in agreement with the General Relativity prediction of GW~\cite{jTjW82,tDjT92}.
Despite this, until very recently there were no direct experimental measurements of GW. %, i.e., they had not previously been observed.

The notion of detecting GW with earth-based experiments began to be devised some fifty years ago \cite{WeberJ}.
%
% Initially resonant mass sensors were proposed, however they did not attain the sensitivity to measure the small deformations produced by GW \cite{}. Subsequently, laser interferometer-based detectors were proposed and by the 2000s an independent global network had been already built, including LIGO \cite{}, Virgo \cite{}, GEO 600 \cite{} and TAMA 300 \cite{}.
%
Then, laser interferometer-based detectors were proposed and by the 2000s an independent global network of detectos such as LIGO \cite{0034-4885-72-7-076901}, Virgo \cite{Virgo2012} and GEO 600 \cite{GEO2014} had been already built. %and TAMA 300 \cite{}.
However, the sufficient enough sensitivity to detect GW was still not reached.
Later, LIGO was subjected to several improvements and upgrades, and on September 12 2015 it becomes the first detector to begin observations with the sufficient sensitive to measure GW from astrophysical origin \cite{0264-9381-32-7-074001}.
Two days later, the advanced LIGO detected the first GW event \cite{Abbott:2016}, and by the end of the year, on December 26 a second GW as also observed \cite{Abbott:2016b}.
These GW signals, named GW150914 and GW151226, were produced by the coalescence of two black holes.
Remarkably, they allowed the direct confirmation of the existence of the GW predicted in the Einstein's theory and also demonstrated the existence of binary stellar-mass systems.
%
% These discoveries have gave new life to the General Relativity \cite{Misner73} providing a new window for astronomy, a new alternative to study and to understand the universe, and the expectation of novel technological developments \cite{sR00}.

% ***********************************
% Planteamiento del problema/necesidad 
% ***********************************
%
It is important to stress that detection of GW150914 and GW151226 was possible due to three fundamental scientific and technological achievements.
First, the solution of the Einstein's equations and the understanding of their physical meaning. For instance, post-Newtonian solutions to the relativistic two body problem and numerical relativity have allowed to obtain accurate GW waveforms during the coalescence of binary black holes.
Second, the LIGO detector, which is an extremely sensitive instrument that is able to measure the very tiny GW signals generated by distant astrophysical systems.
Third, the data analysis algorithms, which aim to pull out GW signals from the measurements provided by LIGO. Indeed, GW signals typically have a lower amplitude than the detector noise, thus data analysis techniques as the matched filter are essential to extract GW.
Given the key role of these three aspects, it is important to have a clear description of them in order to be able to understand GW and its detection.
This work aims to study and to understand the fundamental basis of GW generated by binary systems and to carry out LIGO data analysis to detect this type of GW.
The work contributes in two aspects.
First, it is presented and intuitive and easy-to-understand description of general relativity, GW from binary systems and the basic operation of LIGO.
Second, it is proposed and implemented a methodology based on the matched filter algorithm \cite{PhysRevD.85.122006} to perform detection of GW from binary systems that are embedded in the recorded LIGO data.
The work is presented in a tutorial-like fashion allowing researchers and students interested in the field of GW detection to understand the theoretical concepts and to perform quick basic data analysis devoted to detect GW using interferometric data.
Finally, all the functions implemented in this work (using MATLAB) are freely available at http://www.divulgacienciajalisco.mx/investigacion, thus they can be used to start working with detection of GW signals.

% ***********************************
% Organizacion del trabajo
% ***********************************
%
The manuscript is organized as follows.
Section II describes the fundamental basis of General Relativity, GW and introduces the analytical model of GW generated by inspiral binary systems.
Section III describes the basic principles of LIGO, the analytical model of the strain induced on LIGO by GW from inspiral binary systems and presents the matched filter algorithm. 
Section IV presents a comprehensive data analysis methodology developed to perform detection of GW from binary systems using freely available LIGO data.
The work finishes with conclusions in section V.

%\input{02_GravitationalTheory}
%
%%%%%%%%%%%%%%%%%%%%%%%%%%%%%%%%%%%%%%%
\section{Gravitational Theory}
\label{sec:GravitationalTheory}
%%%%%%%%%%%%%%%%%%%%%%%%%%%%%%%%%%%%%%%

%---------------------------------------
\subsection{Basic description of the General Relativity}
\label{subsec:GeneralRelativity}
%---------------------------------------
%
General Relativity theory describes space and time as part of the same entity, the so-called \textit{spacetime geometry} \cite{Carroll04a}.
According to the theory, the spacetime is affected or curved by massive objects which under non-static conditions can generate ripples that travel at the speed of light.
The simplest representation of this geometry is the Minkowski spacetime, where a line element representing the displacement between two events is expressed by
\begin{equation}
ds^2 = -c^2 dt^2 + dx^2 + dy^2 + dz^2,
\end{equation} 
where, $(t,x,y,z)$ are the coordinates of the geometric system (temporal and spatial coordinates) and $c$ is the speed of light \cite{pRs94}.
To generalize to any geometric system of arbitrary curvature that is described by any coordinate system, the line element is expressed as
\begin{equation}
ds^2 = g_{\mu\nu} dx^{\mu} dx^{\nu},
\end{equation}
where $g_{\mu \nu}$ is the metric tensor, $dx^{\mu}$ and $dx^{\nu}$ are the coordinate directions, and $\mu$ and $\nu$ are indices which run from $0$ to $3$ and represent the time ($0$) and space ($1,2,3$).
Note that the metric tensor $g_{\mu \nu}$ simply defines the relationship between the line element $ds^2$ and the coordinate directions $dx^{\mu}$ and $dx^{\nu}$.

In the General Relativity theory, the geometry of the spacetime is described by ten second-order non-linear differential equations called the \textit{Einstein field equations}
\begin{equation}
G_{\mu\nu} = R_{\mu\nu} - \frac{1}{2}g_{\mu \nu}{\rm R}=  \frac{8\,\pi\,{\rm G}}{c^4}T_{\mu \nu}, \label{equ:EE}
\end{equation} 
where $G_{\mu \nu}$ is the Einstein tensor (i.e., the spacetime geometry), $R_{\mu\nu}$ and $R$ are the Ricci tensor and scalar respectively, $T_{\mu \nu}$ is the stress-energy density tensor (i.e., the matter of an object) and $\rm G$ is the Newton constant of gravitation.
%
% Hence, the Einstein tensor $G_{\mu \nu}$ contains the information of the curvature of a system as well as the lengths described by the metric tensor $g_{\mu \nu}$, while the stress-energy tensor $T_{\mu \nu}$ represents the matter of an object alojado in the spacetime.
%
% The physical significance of these equations is that they link the mass with the curvature it will produce on the spacetime.
%
Thus, more massive objects will produce greater curvature of the spacetime, and consequently, stronger gravitational interaction they will have with other objects.

%---------------------------------------
\subsection{Gravitational Waves}
\label{subsec:GravitationalWaves}
%---------------------------------------
%
Gravitational radiation produced by an object on the spacetime is obtained by the Einstein field equations in the weak field limit \cite{Carroll04a}. 
If the spacetime is flat with a week perturbation on it, we can describe it as
\begin{equation} \label{equ:metricp}
g_{\mu \nu} = \eta_{\mu \nu} + h_{\mu \nu}, \quad h_{\mu \nu} \ll 1,
\end{equation}
where $\eta_{\mu \nu}$ is the flat Minkowski metric and $h_{\mu \nu}$ is a perturbation to the metric. This $g_{\mu \nu}$ is replaced in equation (\ref{equ:EE}) and the linearized Einstein equations are obtained, which must be invariant under coordinate transformation (gauge transformation) \cite{Carroll04a}
\begin{equation}
\overline h_{\mu \nu}=h_{\mu \nu} + 2 \partial_{(\mu} \xi_{\nu)}, \label{gauge}
\end{equation}
where $\xi_{\nu}$ is so small that terms of order $O(h_{\mu \nu}(\partial_\mu \xi^\nu))$ and higher can be neglected of the perturbation.
So, the linearized Einstein field equations become
\begin{equation} \label{equ:pert}
\Box \overline {h}_{\mu \nu}+g_{\mu \nu} \partial^\sigma \partial^\rho \overline h_{\rho \sigma}-\partial^\rho\partial_\nu \overline h_{\mu \rho}-\partial^\rho \partial_\mu \overline h_{\nu \rho} = \frac{-16 \pi{\rm G}}{c^4} T_{\mu \nu}, 
\end{equation}
where $\Box \equiv \partial_\mu\partial^\mu$ is the D'Alambertian operator \footnote{This operator is described by temporal and spacial partial derivatives $\frac{-1}{c^2}\frac{\partial}{\partial \, t^2} + \nabla^2$}.
To further simplify this expression, it is convenient to use harmonic gauge such that $\partial ^{\nu} \overline h_{\mu \nu}=0$, which gives
\begin{equation}
\Box \overline {h}_{\mu \nu} = \frac{-16 \pi{\rm G}}{c^4} T_{\mu \nu}. \label{equ:waveq}
\end{equation} 
If we make a measurement very far from the object, the energy-momentum tensor becomes $T_{\mu \nu}=0$. Substituting the D'Alambertian operator, we obtain the wave equation in vacuum
\begin{equation}
\left( \frac{-1}{c^2}\frac{\partial}{\partial \, t^2} + \nabla^2 \right) \overline h_{\mu \nu} =0,
\end{equation} 
where ${\overline h}_{\mu \nu}$ is the perturbation in the spacetime geometry whose solution is
\begin{equation}
\overline{h}_{\mu \nu}=C_{\mu \nu} \exp (i\kappa_\lambda x^{\lambda}),
\end{equation}
where $C_{\mu \nu}$ is symmetric, transverse-traceless and constant tensor with $C_{0 \mu}=C_{\mu 0}=0$ which contain the polarization of the wave and is called polarization tensor. 
%
% In analogy with classical waves $\kappa^{\lambda}=(\omega, \kappa_1, \kappa_2, \kappa_3)$ are the components of the wave vector. For any non trivial solution, we have $$\kappa_{\lambda}\kappa^{\lambda}=\vec{\kappa}^{2}-{\omega}^2=0,$$ showing that the wave vector is a null vector.
%
This shows that the Einstein field equations lead to an oscillatory activity of the spacetime geometry that travel at the speed of light: \textit{Gravitational Waves (GW)}.

Typical sources of these GW are astrophysical objects such as compact binary systems where each component may be a neutron start (NS) and/or a black hole (BH) \cite{Shapiro}. This objects are highly massive and move at violent accelerations to produce and radiate GW. Additionally, it is believed that the cosmic background radiation also contains gravitational radiation that comes from supernova burst \cite{Supernovae,Supernovae2016} and the big bang \cite{Liddle}.

\begin{figure*}[h]
	\begin{center}
  	\includegraphics[width=0.4\textwidth]{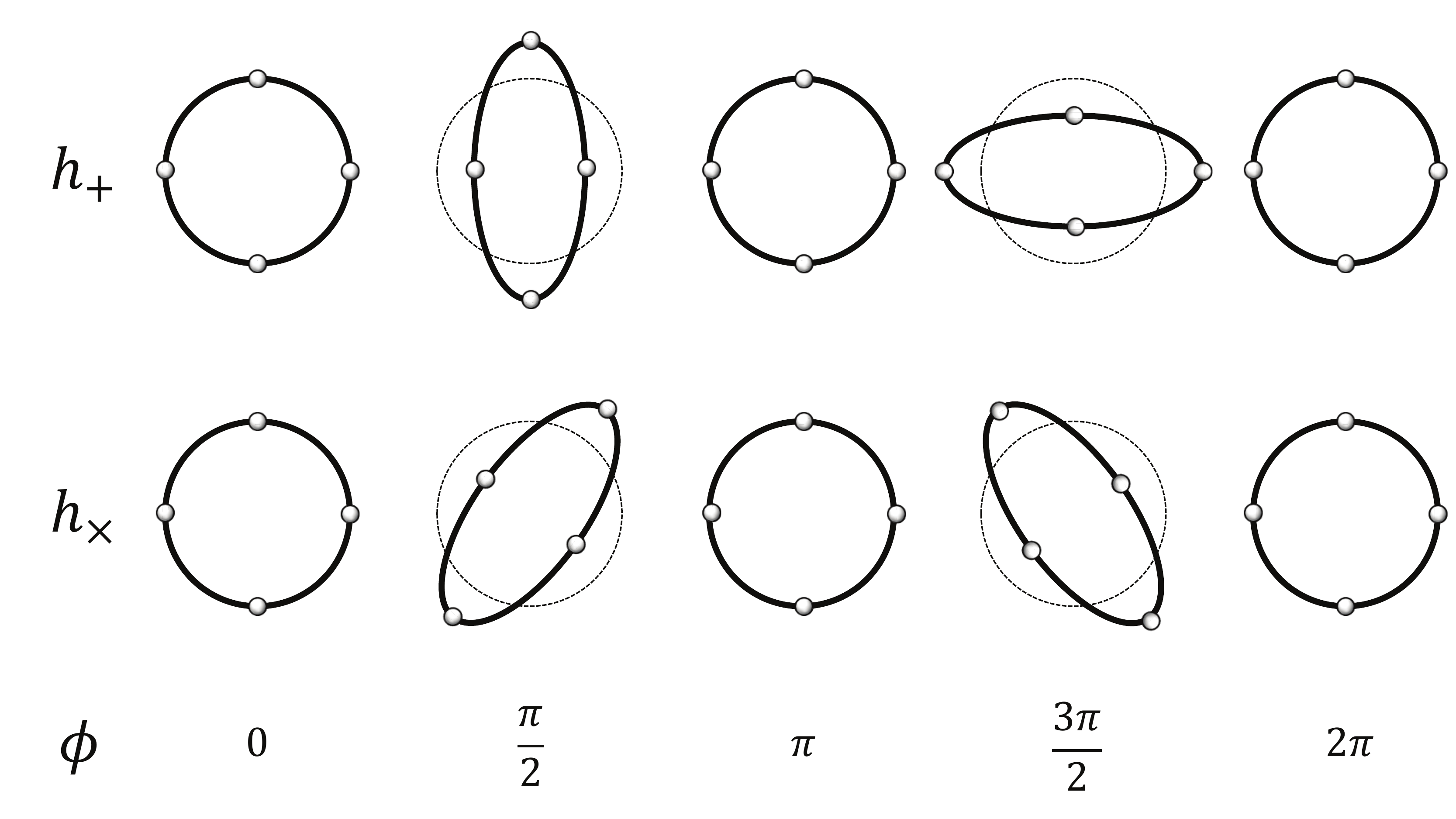}
   	\caption{The physical effect of the GW polarizations $h_+$ and $h_{\times}$. In the illustration, a sinusoidal GW travels through the $z$ axis or perpendicular to the page and the effect over the ring of particle that lay in the $xy$ plane or over the page is to stretch and to squeeze the separation distance between them.}
	\label{fig:polarization}
	\end{center}
\end{figure*}

The spatial components of the GW seen by an observer at a point far from a source, where the source is located at the origin in the $xy$ plane and wave propagates along the $z$-direction (see figure \ref{fig:ReferenceFrameSource} for an illustration) are
\begin{equation}
h^{TT}_{ij}=
\left(\begin{smallmatrix}
h_+ & h_{\times} & 0 \\
h_{\times} & -h_+ & 0 \\
0 & 0 & 0
\end{smallmatrix}\right), \label{trace}
\end{equation}
where $h_+$ and $h_{\times}$ are the \textit{plus} and \textit{cross} independent polarizations \footnote{TT}. Here TT represents Transverse-Traceless tensor because $h_i^i=0$ and $\kappa^\mu C_{\mu \nu}=0$ respectively. 
%From the gauge condition $\overline h_{\mu \nu}=h_{\mu \nu}-\frac{1}{2} \eta_{\mu \nu}h$ and $h=\eta^{\mu \nu} h_{\mu \nu}$ observe that ${\overline h} \equiv \eta^{\mu \nu} {\overline h}_{\mu \nu}=h-2h=-h$, where $\eta_{\mu \nu} \eta^{\mu \nu}=4$, so we can invert the gauge condition $h_{\mu \nu}=\overline h_{\mu \nu}-\frac{1}{2} \eta_{\mu \nu}\overline h$ and ${\overline h}^\mu _\mu=\overline h=0$, gives us ${\overline h}_{\mu \nu}=h_{\mu \nu}$.
%
%
To understand the effect of GW, lets consider a sinusoidal GW propagating through a $z$ axis and a ring of particles that lies in the $xy$ plane (see figure \ref{fig:polarization}).
$h_+$ stretches the ring in the $x$ direction while squeezes it in the $y$ direction for the first half of the cycle and then squeezes the ring in the $x$ direction while stretches it in the $y$ direction for the second half of the cycle.
Similarly, $h_{\times}$ stretches and squeeze the ring but with a gap of $\pi/4$.
%
% To sum up, as the phase of the GW changes, the ring is distorte, in consequence, there is a relative change in the length between of two particles.
%

Thus, the effect of a GW between two separated freely falling particles is to stretch and to squeeze the separation distance between them, i.e., the measured distance $L$ between the particles changes by a distance $\Delta L \sim h L$ where $h$ is the   GW strain.
Thus, $h$ is defined as the fractional change in length between two test masses:
\begin{equation}
h \equiv \frac{\Delta L}{L}.
\end{equation}
%
% However, even the most intense and strongest sources of GW will produce strains in the order of $h \sim 10^{-22}$.

%---------------------------------------
\subsection{Gravitational Waves from Inspiral Binary Systems}
\label{subsec:GravitationalWavesfromInspiralBinarySystems}
\begin{figure*}[h]
	\begin{center}
  	\includegraphics[width=0.5 \textwidth]{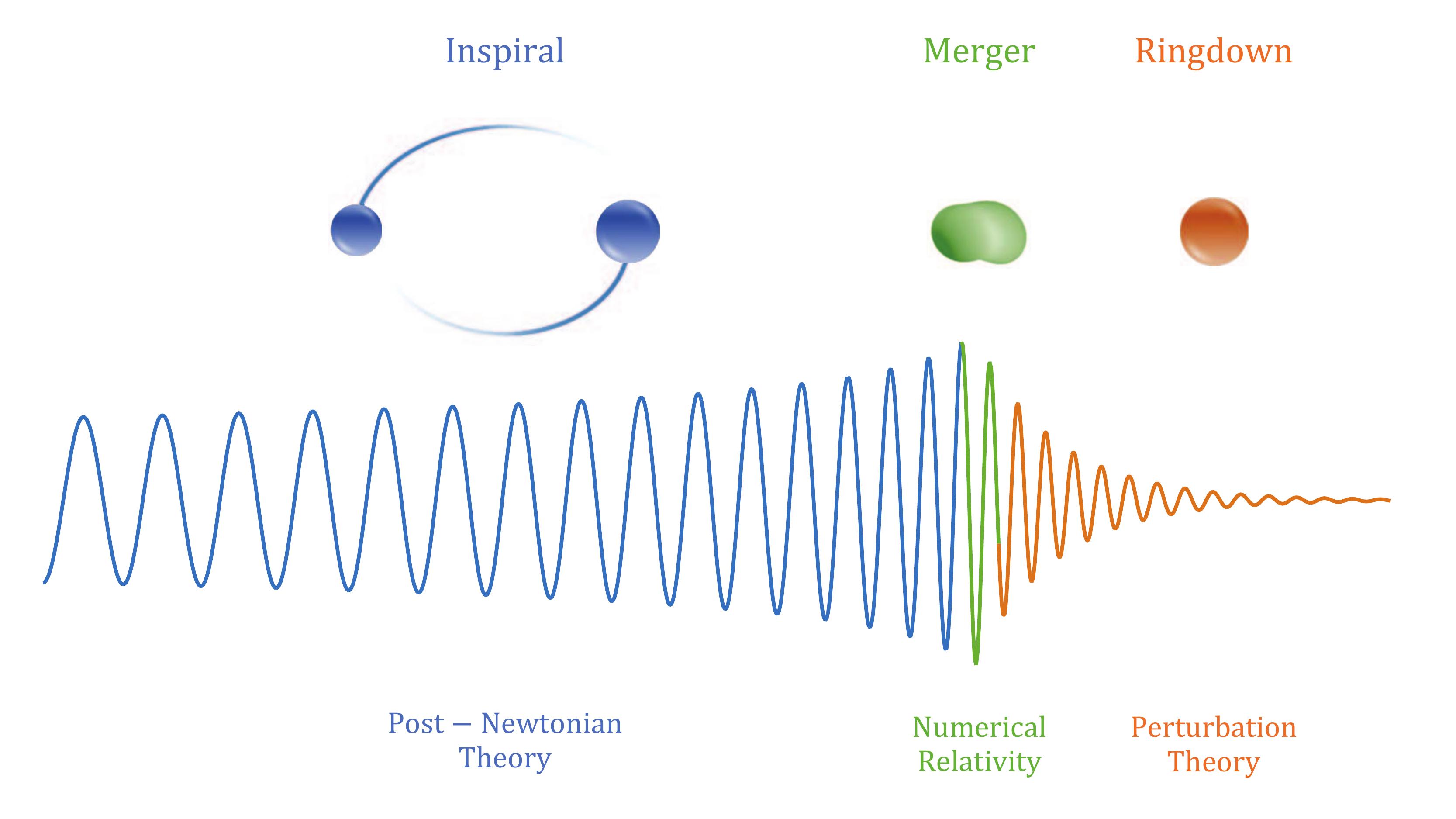}
   	\caption{The three phases in the temporal evolution of a binary system. In the inspiral phase the two object are orbiting and approaching each other. In the merger phase the two objects fuse into one. In the ringdown phase the resulting object relaxes to a stationary state.}
	\label{fig:BinarySystemsPhases}
	\end{center}
\end{figure*}

%---------------------------------------
%
The first two direct observations of GW signals recently reported by the LIGO team (GW150914 and GW151226) were produced by the coalescence of two stellar-mass black holes \cite{Abbott:2016, Abbott:2016b}.
%
% The present work is precisely focused in the study and the data analysis from LIGO to detect GW generated by this type of binary systems.
%
In these systems, the two objects gradually spirals inwards while GW are emitted.
In this process, the system evolves in three different phases, inspiral, merger and ringdown (Figure \ref{fig:BinarySystemsPhases}).

In the \textit{inspiral} phase the two objects are orbiting and approaching each other while the orbital frequency increases. At this stage, post-Newtonian approximations to general relativity analytically model the evolution of the system and thus, high accurate signals can be computed \cite{Blanchet1995}. The resulting GW waveform is a chirp signal, i.e., a sinusoid increasing in frequency and amplitude up to a limit.
The \textit{merger} phase initiates when the separation distance between the two objects reaches the so-called innermost stable circular orbit (ISCO). In consequence, the objects collide and plunge into one. The system is dynamically unstable which leads to a highly complex non-linear system of the Einstein equations where no analytical solution exits. Therefore, numerical relativity is needed to compute the GW signal \cite{0264-9381-26-16-165008}.
Finally, the \textit{ringdown} phase stars after the collision and the resulting object, a black hole, relaxes to a stationary state. In this process, the perturbation theory can used to analytically solve the Einstein equations. This leads to the quasi-normal modes of the final Kerr black hole where the GW signal is described by well-modeled exponentially damped sinusoidal oscillations \cite{PhysRevD.60.022001}.
These three stages in the life of a compact binary system are known as Compact Binary Coalescence (CBC).

%\begin{figure*}[h]
%	\begin{center}
%  	\includegraphics[width=0.45\textwidth]{threestages.jpg}
%   	\caption{Three evolution stages of an binary system.}
%	\label{fig:phases}
%	\end{center}
%\end{figure*}

%$\iota$ is the angle between orbital plane of the source with the observer's line of sight along the $z$-axis
%iota the angle between the orbital angular momentum vector and the line of sight

\begin{figure*}[th]
	\begin{center}
	\begin{tabular}{ c }
  	\includegraphics[width=0.5\textwidth]{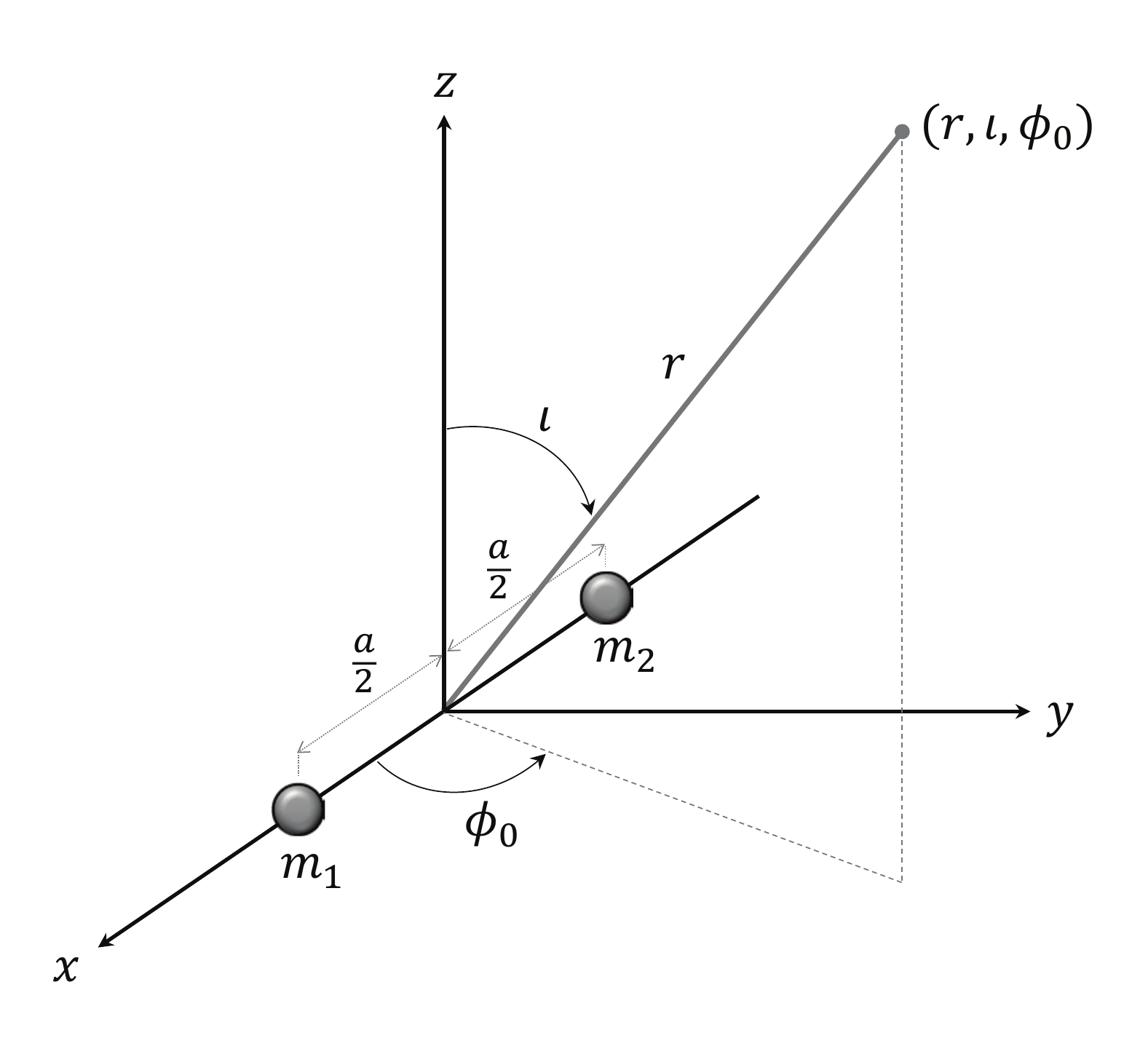} 
	\end{tabular}
  	\\
    %\begin{tabular}{c c}
    %(a) & \hspace{10cm} (b)
    %\end{tabular}
%	\caption{(a) Reference frame of the GW source. The origin is the center of the binary and the GW is observed at point $(r,\iota,\phi_0)$. (b) Reference frame of an interferometer-based detector. The origin is at the detector and the binary system is located at point $(r,\theta,\varphi)$.}
	\caption{Reference frame of the GW source. The origin is the center of the binary and the GW is observed at point $(r,\iota,\phi_0)$.}
	\label{fig:ReferenceFrameSource}
	\end{center}
\end{figure*}

This work focuses in the detection of GW from the inspiral phase using LIGO data.
To accomplish this, we need the analytical model of the GW. 
% and of the GW strain produced in an interferometer detector.
%
%
% The relativistic two-body problem, i.e. the problem of describing the dynamics and gravitational radiation of two extended bodies interacting gravitationally according to general relativity theory, is very difficult.
%
% Figure \ref{fig:ReferenceFrameSource} shown the reference frame for the source, i.e., the binary system.
%
The masses of the two astrophysical objects are $m_1$ and $m_2$, they are separated a distance $a$ and are orbiting in their common center of mass.
The reference frame of the source (Figure \ref{fig:ReferenceFrameSource}) is the Cartesian coordinate system $(x,y,z)$ where the origin is the center of the binary, and the GW is observed at point $(r,\iota,\phi_0)$ where $r$ is the separation distance between the source and the observer, $\iota$ is the angle between the source's angular momentum axis and the observer's line of sight or simply the inclination angle, and $\phi_0$ is the angle between the binary system and the fix reference frame of the detector or simply the orbital phase of the binary. %(see figure \ref{fig:ReferenceFrames}a).
%
%The reference frame of an interferometer-based detector is the Cartesian coordinate system $(x',y',z')$ where the arms are located in the $x'$ and $y'$ axis (see figure \ref{fig:ReferenceFrames}b). Therefore, the source is at point $(r,\theta,\varphi)$ where $\theta$ and $\varphi$ are the sky coordinates of the binary.
%
%To account for the inclination of the angular momentum of the binary system, the phase $\psi$ is employed, which represents the polarization angle of the binary (not shown in figure \ref{fig:ReferenceFrames}) \MA{cambiar la forma de esta tabla de variables}.
%%
%In summary, a total number of nine parameters are needed to describe the GW and the GW strain. These parameters are:
%%
%\begin{equation*}
%\begin{split}
%m_{1},m_{2}      &\quad \textrm{masses of the two astrophysical objects}  \\
%r                &\quad \textrm{the distance to the binary system}        \\
%\iota            &\quad \textrm{inclination angle of the binary system}   \\
%\phi_0           &\quad \textrm{oribital phase of the binary system}      \\
%t_c              &\quad \textrm{end time of the inspiral}                 \\
%(\theta,\varphi) &\quad \textrm{sky coordinates of the binary system}     \\
%\psi             &\quad \textrm{polarization angle of the binary system}
%\end{split}
%\end{equation*}
%%
%But if we consider the spin of the objects, we need to incorporate the magnitude and the two orientation angles for each of them, thus a total of fifteen parameters are needed \cite{}.
%%
%In this work, we will consider the simplest case of non-spinning objects.
%

%---------------------------------------
\subsection{Gravitational Wave in the inspiral phase: $h_{+}(t)$ and $h_{\times}(t)$}
\label{subsec:Gravitationalwaveintheinspiralphase}
%---------------------------------------
%
Considering the reference frame of the GW source, the temporal evolution of the $+$ and $\times$ polarizations of the GW observed at point $(r,\iota,\phi_0)$ are \cite{Maggiore2007,Sathyaprakash:2009xs}
\begin{align}
h_{+}(t) &= \frac{4}{r} \left( \frac{G\mathcal{M}}{c^2} \right)^{5/3} \left( \frac{\pi f_{gw}(t)}{c} \right)^{2/3} \frac{1 + \cos^2\iota}{2} \cos \phi_{gw}(t),
\\
h_{\times}(t) &= \frac{4}{r} \left( \frac{G\mathcal{M}}{c^2} \right)^{5/3} \left( \frac{\pi f_{gw}(t)}{c} \right)^{2/3}  \cos\iota  \sin \phi_{gw}(t),
\end{align}
where $\mathcal{M}=\mu^{3/5}M^{2/5}$ is the chirp mass, $\mu=m_1m_2/M$ is the reduced mass, $M=m_1+m_2$ is the total mass, and $\phi_{gw}(t)$ and $f_{gw}(t)$ are the phase and frequency of the GW, respectively.

To compute the phase and the frequency, the simplest case is to assume that the binary system evolves through a sequence of quasi-stationary circular orbits (i.e., circular binary system).
In this case, the classical Newtonian order solution for $\phi_{gw}(t)$ and $f_{gw}(t)$ leads to
\begin{align} \label{equ:GWPhaseNewtonian}
\phi_{gw}(t) &= - \frac{2}{5} \frac{c^3}{GM} \left[\Theta(t)\right]^{-\frac{3}{8}} \left( t_c - t\right) + \phi_c,
\\  \label{equ:GWFrequencyNewtonian}
f_{gw}(t) &= \frac{c^3}{8\pi GM} \left[\Theta(t)\right]^{-\frac{3}{8}},
\end{align}
where $\Theta(t) = (c^3\eta/5GM)(t_c - t)$ is a dimensionless time-dependent variable, $\eta=\mu/M$ is the ratio between the reduced mass $\mu$ and the total mass $M$, $t_c$ is the end time of the inspiral or the coalescence time and $\phi_c$ is the value of the orbital phase at $t_c$.  % innermost stable circular orbit
%
% Note that $f(t) = 2 \frac{1}{2\pi} \frac{d \phi(t)}{dt}$, where a factor of 2 was used due to the fact that $\phi(t)$ (i.e., the phase of the GW) is twice the phase of the orbital motion, therefore, the frequency of the GW is twice the orbital frequency).
%
% Figure \ref{fig:GWwaveform} shows the waveform of the $+$ and $\times$ polarization of a GW emitted from an inspiral binary system with masses $m_1=10M_\odot$ and $m_2=10M_\odot$ observed at a distance $r=100Mpc$ with inclination angle $\iota=0$ and orbital phase $\phi_0=0$.
%
Note that as the time increases (i.e., the end of the inspiral phase is approached) the frequency and amplitude of the GW increase, therefore the GW signal $h_{+}(t)$ and $h_{\times}(t)$ exhibit the so-called \emph{chirp} type waveform.

It is important to note that this Newtonian order solution is not sufficient to correctly model the GW waveform.
However, precise well-models are relevant to carry out successful detection of GW.
Hence, post-Newtonian (PN) order corrections are necessary.
%
% Roughly, the PN correction provides a more accurate solution for the motion, the rates of emission of energy and the angular momentum of the binary system.
%
These corrections are embodied in the form of power series of the ratio $v/c$, where $v$ is the orbital velocity of the binary system.
%
% Note however that PN correction is valid if the objects of the binary can be modeled as point particles (no se que aporta esto, pero lo lei en algun lado).
%
The post-Newtonian waveform models the amplitude evolution using the quadrupole formula, but includes higher-order terms corrections to the phase.
The phase and frequency time-evolution of the GW in the second-order post-Newtonian (2PN) approximation of general relativity are \cite{Blanchet1996}
\begin{eqnarray}
\phi_{gw}(t) &=& \phi_0 - \frac{1}{\eta} \left[ 
\Theta(t)^\frac{5}{8} + \left(\frac{3715}{8064} + \frac{55}{96}\eta\right)
\Theta(t)^\frac{3}{8} - \frac{3\pi}{4} \Theta(t)^\frac{1}{4} 
\right. \nonumber \\ &&+ \left. \left(\frac{9\,275\,495}{14\,450\,688} + \frac{284\,875}{258\,048}\eta +
\frac{1855}{2048} \eta^2\right) \Theta(t)^\frac{1}{8} \right],
\end{eqnarray}
\begin{eqnarray}
f_{gw}(t) &=& \frac{c^3}{8\pi GM} \left[ \Theta(t)^{-\frac{3}{8}} + \left({743 \over 2688} + {11 \over 32}\eta \right) \Theta(t)^{-\frac{5}{8}}  - {3\pi \over 10} \Theta(t)^{-\frac{3}{4}}  
\right. \nonumber \\ &+& \left. \left( {1855099\over 14450688}+{56975\over 258048} \eta + {371\over 2048} \eta^2 \right) \Theta(t)^{-\frac{7}{8}} \right].
\end{eqnarray}
Note that these expressions are extensions (i.e., the include correction terms) for the phase and frequency of the GW in the classical Newtonian order solution presented in equations (\ref{equ:GWPhaseNewtonian}) and (\ref{equ:GWFrequencyNewtonian}).

%\input{03_DetectionOfGW}
%
%%%%%%%%%%%%%%%%%%%%%%%%%%%%%%%%%%%%%%%
\section{Detection of Gravitational Waves}
\label{sec:DetectionGravitationalWaves}
%%%%%%%%%%%%%%%%%%%%%%%%%%%%%%%%%%%%%%%

%---------------------------------------
\subsection{The laser interferometer gravitational-wave observatory (LIGO)}
\label{subsec:LIGO}
%---------------------------------------
%
GW generated by astrophysical objects will produce on Earth strengths in the order of $h \sim 10^{-22}$. To measure these tiny strain levels, it is needed a highly sensitive instrument able to detect relative changes of distance of $\sim 10^{-22}$ times the reference length $L$ of the instrument. The laser-based interferometer is the scientific experiment that can perform these very small measurements. Interferometers were proposed in 1960's to measure relative changes in length produced by GW \cite{WeberJ, Weber:1961}, and the first prototype was constructed and tested by 1970's \cite{Forward:1971, Forward:1978}. Professors Rainer Weiss from MIT and Kip Thorne from Caltech devised the fundamental design of an interferometer-based detector with adequate characteristics to reach the required sensibility to measure GW \cite{Weiss:1972,Drever:1981, Drever:1989}. This was the birth of the Laser Interferometer Gravitational-Wave Observatory (LIGO), an extraordinary scientific experiment devoted to measure changes in length even smaller than the diameter of the proton that are produced by gravitational sources of astrophysical origin  \cite{0034-4885-72-7-076901}. LIGO consists of two L-shape power-recycling Fabry-Perot Michelson interferometers with arms-length of 4 Km \cite{0264-9381-32-7-074001,0264-9381-27-8-084006}. The interferometers are located in Hanford, WA (referred as H1) and in Livingston, LA (referred as L1), and they are separated approximately 3000 Km, which is equivalent to about 10 ms at the speed of light. Although the interferometers are almost identical (the only difference is the position on earth), at least two are required in separate places in order to validate and confirm GW events by coincident detections \cite{0264-9381-28-12-125023,PhysRevD.88.024025} and to estimate the position of the source in the sky.

\begin{figure*}[h]
	\begin{center}
  	\includegraphics[width=0.60\textwidth]{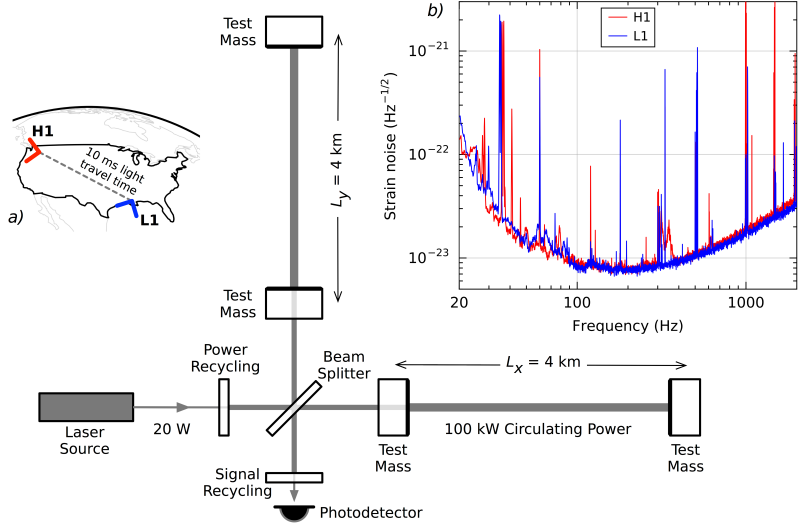}
   	\caption{Illustration of the basic LIGO interferometer with its main components: the test masses, the Fabry-Perot cavities and power-recycling mirror. (a) Location of the two LIGO interferometers on earth. The separation distance is 3002 Km or 10 ms of light travel time. (b) Sensibility curves of the two LIGO interferometers. Image from \cite{Abbott:2016}.} % (Source: https://www.ligo.caltech.edu)
	\label{fig:LIGO}
	\end{center}
\end{figure*}

To illustrate the basic principle of LIGO, a simple Michelson interferometer with Fabry-Perot and power-recycling systems is illustrated in figure \ref{fig:LIGO}. In operation, the laser light is guided towards a beam splitter which reflects half of the light into the $x$-arm and half of the light into the $y$-arm. Both light beams present a phase shift of half a cycle. The light travels in each arm up to the end mirrors and is reflected back towards the beam splitter where they are summed. The resulted laser light is then collected with a photodiode. If no GW arrives, the two arms remain at the same original length and the two laser lights cancel out each other (destructive interference). Thus, no GW strain signal is observed. If a GW arrives, the length of the arms changes. This leads to different time traveling of the light in the arms and to a phase shift different to half a cycle. In consequence, the two laser lights add up (constructive interference) and a GW strain signal is observed. The Fabry-Perot system consist of mirrors placed in the arms near to the beam splitter. These inner mirrors are designed to keep the light bouncing several times to enlarge the phase shift to measurable levels. The power-recycling system aims to increase the effective laser power in the arms without increasing the power of the laser source. This system is based on a mirror, located between the beam splitter and the laser source, that reflects the laser light (received from the arms) back to the beam splitter. 

A LIGO inteferometer is able to measure length changes in the order of $\Delta L = h \times L \sim 10^{-19}$ m. This remarkable sensitivity is essential to measure GW from typical sources of astrophysical origin, however it also leads to the observation of noise from several unwanted sources \cite{0034-4885-72-7-076901}. Fundamental noise sources are $(i)$ seismic noise, $(ii)$ suspension thermal noise, and $(iii)$ photon shot noise. Thus, the recorded signal from LIGO is
\begin{equation} \label{equ:LIGOsignal2}
s(t) = w(t) + h(t),
\end{equation}
where $w(t)$ is the noise and $h(t)$ is the GW strain (only present if a GW is arriving to the detector). The noise is always present and defines the sensibility of LIGO, which is characterized by the amplitude spectral density, i.e., the square root of the power spectral density (PSD) of the data recorded in the absence of GW signals. LIGO was improved until it reached sensitivity level of $\sim 10^{-22}$ at $\sim$ 100 Hz \cite{SensitivityLIGO}. Figure \ref{fig:LIGO} shows the attained sensibility in the two LIGO interferometers (H1 and L1) that finally allowed the detection of the recently reported GW events \cite{0264-9381-33-13-134001}.

%---------------------------------------
\subsection{Gravitational Wave strain induced in LIGO}
\label{subsec:StrainInducedInLIGO}
\begin{figure*}[th]
	\begin{center}
	\begin{tabular}{ c }
  	\includegraphics[width=0.5\textwidth]{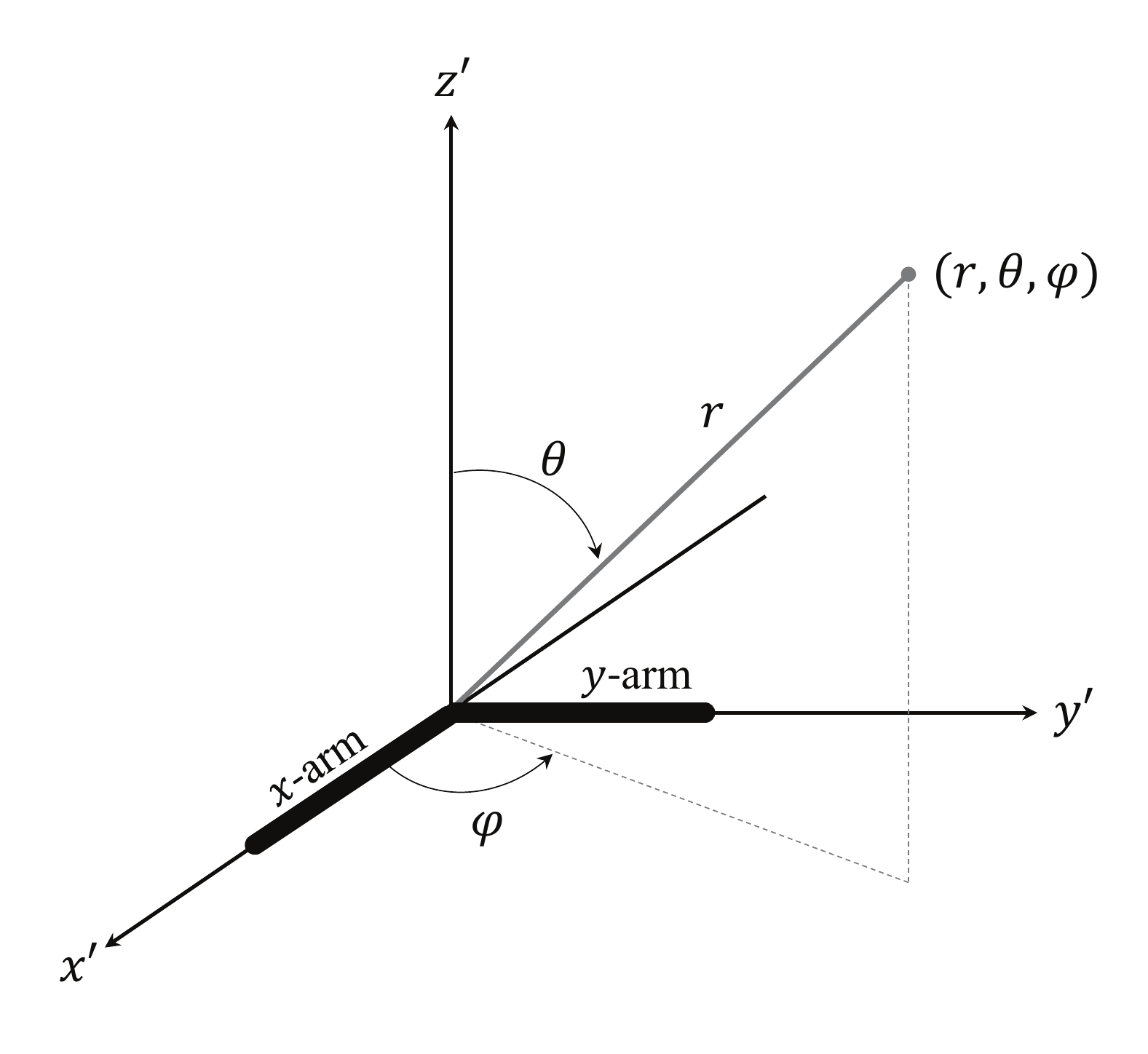} 
	\end{tabular}
	\caption{Reference frame of an interferometer-based detector. The origin is at the detector and the binary system is located at point $(r,\theta,\varphi)$.}
	\label{fig:ReferenceFrameDetector}
	\end{center}
\end{figure*}

%---------------------------------------
%
To detect GW using LIGO data, we need the analytical model of the GW strain induced in the interferometer detector. Such reference frame is the Cartesian coordinate system $(x',y',z')$ where the arms are located in the $x'$ and $y'$ axis (Figure \ref{fig:ReferenceFrameDetector}). Therefore, the source is at point $(r,\theta,\varphi)$ where $\theta$ and $\varphi$ are the sky coordinates of the binary.

\subsubsection{Time-domain strain signal: $h(t)$}
Considering the reference frame of an interferometer-based detector, the GW strain produced by a source located at $(r,\theta,\varphi)$ is a linear superposition of the $+$ and $\times$ polarizations of the GW according to the detector's response \cite{0264-9381-28-12-125023}
\begin{equation} \label{equ:TimeDomainStrain1}
h(t) = F_{+} h_{+}(t) + F_{\times} h_{\times}(t), 
\end{equation}
where $F_{+}$ and $F_{\times}$ are the beam-pattern functions that model the amplitude response of the detector to the $+$ and $\times$ polarizations and depend on geometrical factors ($\theta$, $\varphi$ and $\psi$). Therefore, the GW strain induced in an interferometer-based detector is
\begin{equation} \label{equ:TimeDomainStrain2}
h(t) = \frac{A(t)}{\mathcal{D}} \cos\left( 2 \phi_{gw}(t) - \theta \right),
\end{equation}
where $A(t) = - (2G\mu / c^4 ) \left( \pi GM f_{gw}(t) \right)^\frac{2}{3}$ is the time-dependent quadrupolar amplitude, $\theta$ is a phase angle comprising the inclination angle and detector's response $\tan \theta = \frac{F_\times}{F_+} \frac{\cos \iota}{(1 + \cos^2 \iota)/2}$ and $\mathcal{D}$ is the \textit{effective distance} of the source
\begin{equation}
\mathcal{D} = \frac{r}{\sqrt{F_+^2 \left( \frac{1 + \cos^2 \iota}{2} \right) ^2 + F_\times^2 \cos^2 \iota}}.
\end{equation}
Note that the effective distance $\mathcal{D}$ is related to real distance $r$ by the geometrical parameters that relate the source orientation to the detector orientation. Thus, $\mathcal{D}$ is equal to $r$ when the source is \emph{optimally-oriented} (the inclination angle is $\iota = 0$, i.e., the source is located on the $z$-axis on the observer's line of sight; and the source location is at sky position $\theta = 0$ or $\pi/2$, i.e. above or below the zenith of the detector), and $\mathcal{D}$ is greater than $r$ when the source is \emph{sub-optimally-oriented}. 

\subsubsection{Frequency-domain strain signal: $\tilde{h}(f)$}
To carry out GW detection from inspiral binary systems using LIGO data, the frequency domain representation of the GW strain is required. The Fourier transform of $h(t)$ could be directly applied to compute $\tilde{h}(f)$, however, this is computationally expensive. The stationary phase approximation is an alternative efficient method to express chirp-type waveforms, as GW signals, directly in the frequency domain \cite{Droz:1999qx}. The GW strain in the frequency-domain using this method is given by
\begin{equation} \label{equ:FrequencyDomainStrain}
  \tilde{h}(f)   = \left( \frac{1}{\mathcal{D}} \right)  \, \, {\mathcal{A}}_{} \, \, f^{-7/6}  \, \, \ensuremath{\mathrm{e}}^{-i\Psi(f)} ,
\end{equation}
where ${\mathcal{A}}_{}$ is an amplitude term constant in frequency that is parametrized only by the masses of the source
\begin{equation}
  {\mathcal{A}}_{} = -\left(\frac{5}{24\pi}\right)^{1/2} \left(\frac{GM_\odot}{c^2}\right) \left(\frac{\pi GM_\odot}{c^3}\right)^{-1/6} \left(\frac{\mathcal{M}}{M_\odot}\right)^{-5/6} ,
\end{equation}
and $\Psi(f)$ is the phase of term, which in the 2PN order correction is
\begin{equation}
\begin{split}
\Psi(f) &= 2\pi ft_c-2\phi_0-\pi/4+\frac{3}{128\eta}\biggl[v^{-5}+\left(\frac{3715}{756}+\frac{55}{9}\eta\right)v^{-3}
\\
&\quad -16\pi v^{-2}+\left(\frac{15\,293\,365}{508\,032}+\frac{27\,145}{504}\eta+\frac{3085}{72}\eta^2\right)v^{-1}\biggr] ,
\end{split}
\end{equation}
where $v=(\pi M f G/c^3)^{1/3}$. Note that the frequency content of the observed GW strain is limited by the sensitivity frequency band of the detector (see in figure \ref{fig:LIGO}b the reduced sensitivity at low and high frequencies). For the case of LIGO, the expression of $\Psi(f)$ is valid for a frequency range $[ f_{\text{low}} \, f_{\text{high}} ]$, where $f_{\text{low}} = 40$~Hz is defined by the frequency response of the detector and $f_{\text{high}}$ is defined by frequency at the end of the chirp which can be approximated in the ISCO limit as $f_{\text{ISCO}} = c^3 / (6\sqrt{6}\pi GM)$.

%---------------------------------------
\subsection{Matched Filter}
\label{subsec:matchedFilter}
%---------------------------------------
%
Let $s(n)=[s(0),s(1),...,s(N-1)]^T$ (or simply $\textbf{s}$ in vector notation) be a discrete time observation from LIGO, which may or may not contain a GW strain signal $h(n)=[h(0),h(1),...,h(N-1)]^T$ (or simply $\textbf{h}$) whose waveform is known in advance (if we are searching a GW from a binary system with masses $m_1$ and $m_2$ located at a distance $\mathcal{D}$, we can readily computed the time domain GW strain waveform using equation \ref{equ:TimeDomainStrain2}). The goal is to determine if $s(n)$ is only noise or if it contains the waveform $h(n)$ embedded in the noise. This is a classical problem of the detection theory for which we need to design a decision rule for selecting one of two mutually exclusive hypotheses
\begin{equation}
\begin{cases}
\mathcal{H}_0:  s(n) = w(n)        & \text{GW signal is absent},
\\
\mathcal{H}_1:  s(n) = w(n) + h(n) & \text{GW signal is present},
\end{cases}
\end{equation}
where $w(n)$ is the noise in the detector which is modelled with a probability distribution function (PDF). Hence, the PDF of the observed data under the null hypothesis $\mathcal{H}_0$ is $\textbf{s} \sim p(\textbf{s};\mathcal{H}_0)$, while under the alternative hypothesis $\mathcal{H}_1$ is $\textbf{s} \sim p(\textbf{s};\mathcal{H}_1)$. Accordingly, any decision rule will lead to false alarms ($P_{FA}$ is the probability of deciding $\mathcal{H}_1$ when $\mathcal{H}_0$ is true) and false dismissals ($P_{FD}$ is the probability of deciding $\mathcal{H}_0$ when $\mathcal{H}_1$ is true). Under these situations, the optimal decision rule can be achieved with the Neyman-Pearson criterion which minimizes $P_{FD}$ (equivalently maximizes the probability of detection or $P_{D}=1-P_{FD}$) subject to a given $P_{FA}$. Therefore, we decide $\mathcal{H}_1$ if the likelihood ratio test exceeds a threshold, or we decide $\mathcal{H}_0$ otherwise \cite{Kay93b}
\begin{equation}
\frac{p(\textbf{s};\mathcal{H}_1)}{p(\textbf{s};\mathcal{H}_0)}
\begin{matrix}
\mathcal{H}_{1}\\
>\\
<\\
\mathcal{H}_{0}%
\end{matrix}
\gamma
\end{equation}
where the threshold $\gamma$ is chosen to satisfy a given $P_{FA}$. If the noise is white and Gaussian (WGN), this framework leads to the \emph{matched filter} detector \cite{HELSTROM1968102}, which simply correlates the observed data $s(n)$ with the known waveform $h(n)$. Then, the GW strain signal $h(n)$ is present in the observed data $s(n)$ if the test statistics $T(\textbf{s}) = \sum_{n=0}^{N-1} s(n)h(n) = \textbf{s}^{T}\textbf{h}$ is greater than a threshold imposed for a given $P_{FA}$, otherwise, the GW strain signal is not present in the observed data.

The matched filter is the optimal decision rule in the sense that maximises the signal-to-noise ratio (SNR) at the output. However, no detection will be achieved if the expected signal $h(n)$ does not begin at the initial time $n=0$. This is the case with LIGO data as the length of $s(n)$ (several seconds or minutes) is much larger than the length of $h(n)$ (from milliseconds up to a few seconds according to the duration of the chirp), and we do not know in advance at which time the expected GW signal arrives. Therefore, it is required to search efficiently in all times within the observed data, which can be carried out by computing the matched filter in the frequency domain. Indeed, when the noise is wide-sense stationary (WSS) and the observed data length is large, assumptions that hold for LIGO detectors in the periods of time where the search of GW is performed \footnote{It is important to note that the noise in LIGO is non-stationary, in consequence, the noise PSD changes across time. Nonetheless, it is still possible to approximate the noise as stationary over the time periods where GW are searched and to compute a representative noise PSD for that interval.}, the matched filter output is computed as \cite{Kay93b,HELSTROM1968102}
\begin{equation} \label{equ:matchedfilter1}
z(n) = 4 \sum_{k=1}^{\frac{N-1}{2}} \frac{s(f_k) h^\ast(f_k)}{P_{ww}(f_k)} e^{i 2 \pi n k /N},
\end{equation}
where $s(f_k)$ and $h(f_k)$ are the discrete Fourier transform of $s(n)$ and $h(n)$ respectively, $P_{ww}(f_k)$ is the one-sided power spectral density (PSD) of the noise (i.e., LIGO sensitivity) and $^\ast$ denotes complex conjugate. Basically, the matched filter consists in the frequency domain correlation between the observed strain data and the theoretical model of the expected GW strain, with the noise PSD providing a weighting effect where frequency bands of more importance are those for which the noise is small. This is a very important feature for the detection of GW because it addresses the highly frequency-dependent behavior of the LIGO sensitivity and the effect of the spectral peaks (see figure \ref{fig:LIGO}). Note that $s(f_k)$ and $P_{ww}(f_k)$ can be efficiently and readily computed via the fast Fourier transform (\textit{FFT}) algorithm, while $h(f_k)$ is the discrete version of the frequency domain GW strain $\tilde{h}(f)$ presented in equation \ref{equ:FrequencyDomainStrain}.

Observe that $z(n)$ is a complex time series where $\Re\{z(n)\}$ and $\Im\{z(n)\}$ represent the filter output for the known GW signal with a phase of $0$ and $\pi/2$, respectively. Hence, the amplitude SNR is: %Observe that $z(n)$ is a complex function, hence the amplitude SNR is
\begin{equation}  \label{equ:matchedfilter2}
\rho(n) = \frac{|z(n)|}{\rho_{h}},
\end{equation}
where $\rho_{h}$ is a normalization constant representing the sensitivity of LIGO for the known GW (i.e., the SNR that the expected GW signal would attain in the LIGO detector or $SNR_{expected}$) and is given by
\begin{equation} \label{equ:matchedfiltersigma}
\rho_{h} = 2 \sqrt{ \Delta f \sum_{k=1}^{\frac{N-1}{2}} \frac{|h(f_k)|^2}{P_{ww}(f_k)} },
\end{equation}
where $\Delta f$ is simply the frequency resolution. Notice that larger values of $\rho_{h}$ correspond to a highly sensitive detector while smaller values correspond to a noisier detector (due to the effect of the $1/P_{ww}(f_k)$ term).

Finally, detection of GW signals involves: $(i)$ selection of peaks in $\rho(n)$ as larger values suggest that the expected GW strain signal best matches the observed strain data; $(ii)$ a significant test to reject spurious detections that do not represent GW \cite{PhysRevD.71.062001}; and $(iii)$ the estimation of the time instant for the final of the inspiral phase or coalescence time $\widehat{t_c}$ and the corresponding recovered SNR ($SNR_{recovered}$). It is important to emphasize that in the process of searching GW using LIGO data, the parameters of the source (such as the masses and the distance) are not known in advance, but the GW strain required for the matched filter depends upon those parameters. Therefore, a set of GW waveforms that cover the parameters space, called template bank, has to be generated and the search is carry out for each of those waveforms \cite{0264-9381-23-18-002,0264-9381-24-24-006}.

This fundamental data analysis procedure was described in the \emph{FindChirp algorithm} \cite{PhysRevD.85.122006,Brown:2004vh} for the detection of GW form inspiral binary systems \cite{PhysRevD.87.024033}, and has been essential for the first two detections of GW recently reported by the LIGO Scientific Collaboration \cite{Abbott:2016,Abbott:2016b,PyCBC2016,PhysRevD.90.082004}.

%\input{04_DataAnalysis}
%
%%%%%%%%%%%%%%%%%%%%%%%%%%%%%%%%%%%%%%%
\section{Data Analysis}
\label{sec:DataAnalysis}
%%%%%%%%%%%%%%%%%%%%%%%%%%%%%%%%%%%%%%%
%
This section describes the methodology proposed to search and to assess detection of GW waveforms from binary systems using freely available data from LIGO.

%---------------------------------------
\subsection{Dataset description}
\label{subsec:DatasetDescription}
%---------------------------------------
%
Strain data from LIGO have been recorded during several observing periods called science runs. These recordings periods have been carried out to validate, calibrate and improve all the systems as well as to search for GW. The LIGO Open Science Center (LOSC) provides free access to these data records with the aim to be used and studied by anyone interested in the field of GW detection \cite{1742-6596-610-1-012021,PhysRevD.82.102001}.

The fifth science run (S5) was carried out when the LIGO detector finally reached the initial design sensitivity \cite{2010NIMPA.624..223A,0264-9381-23-19-S03}. Data from S5 \cite{LIGOSC2014} was released on 22 August 2014 and the recording lasted almost two years (from 4 November 2005 until 1 October 2007). This data was subjected to the search of GW from different sources of astrophysical origin but none were detected \cite{PhysRevD.79.122001,PhysRevD.80.047101,PhysRevD.83.122005}. Interestingly, the recorded data contain the strain of simulated but realistic GW, which were injected in both H1 and L1 detectors by carefully and adequately moving the arm's mirrors. All injections of GW from inspiral binary systems were based on the second-order post-Newtonian correction to compute the GW waveform (see subsection \ref{subsec:StrainInducedInLIGO}). In the present work, this LIGO data was used to examine detection of GW from inspiral binary systems. 

\begin{figure*}[h]
	\begin{center}
	\begin{tabular}{ c c }
  	\includegraphics[width=0.5\textwidth]{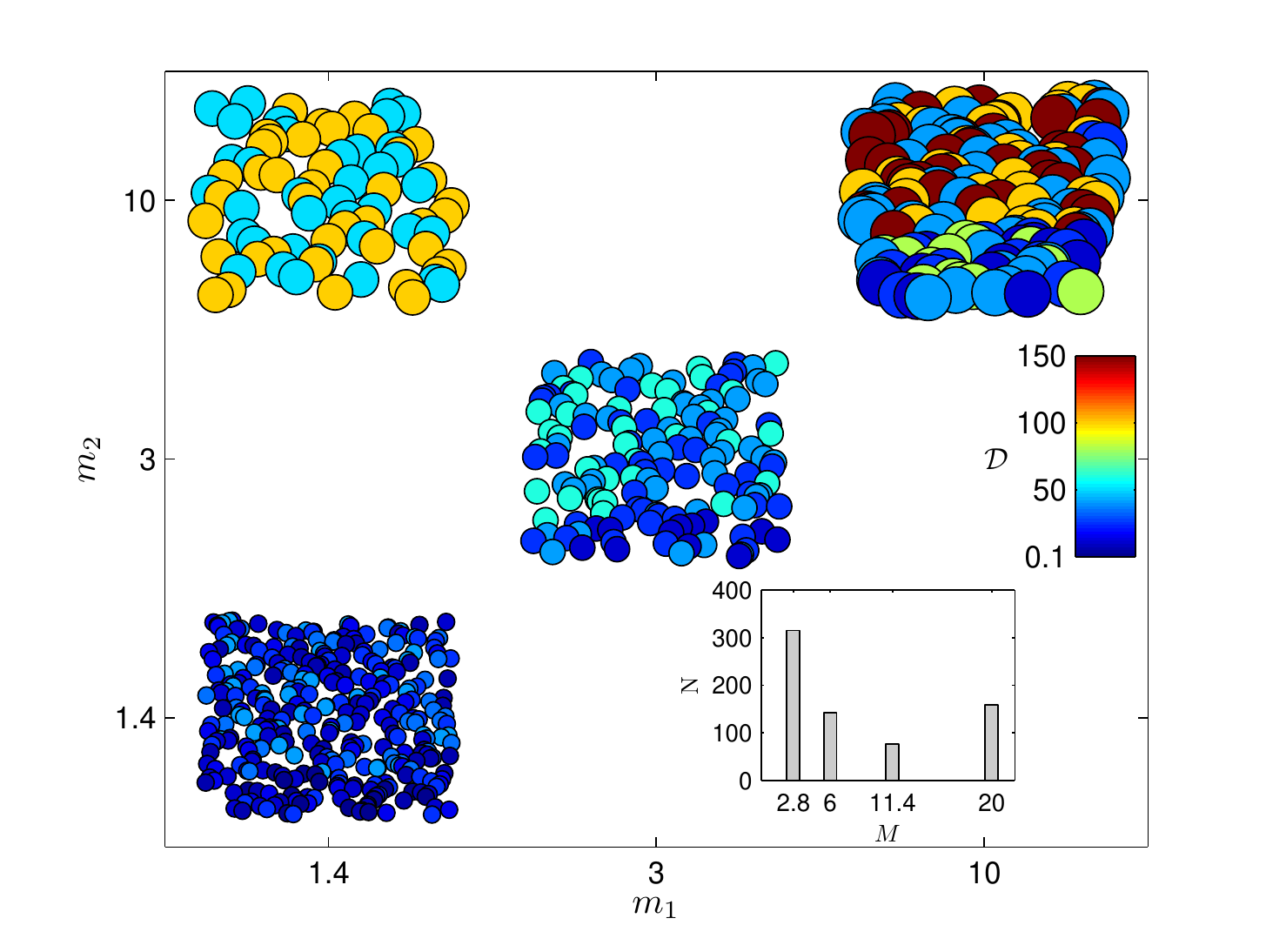} 
  	&
  	\includegraphics[width=0.5\textwidth]{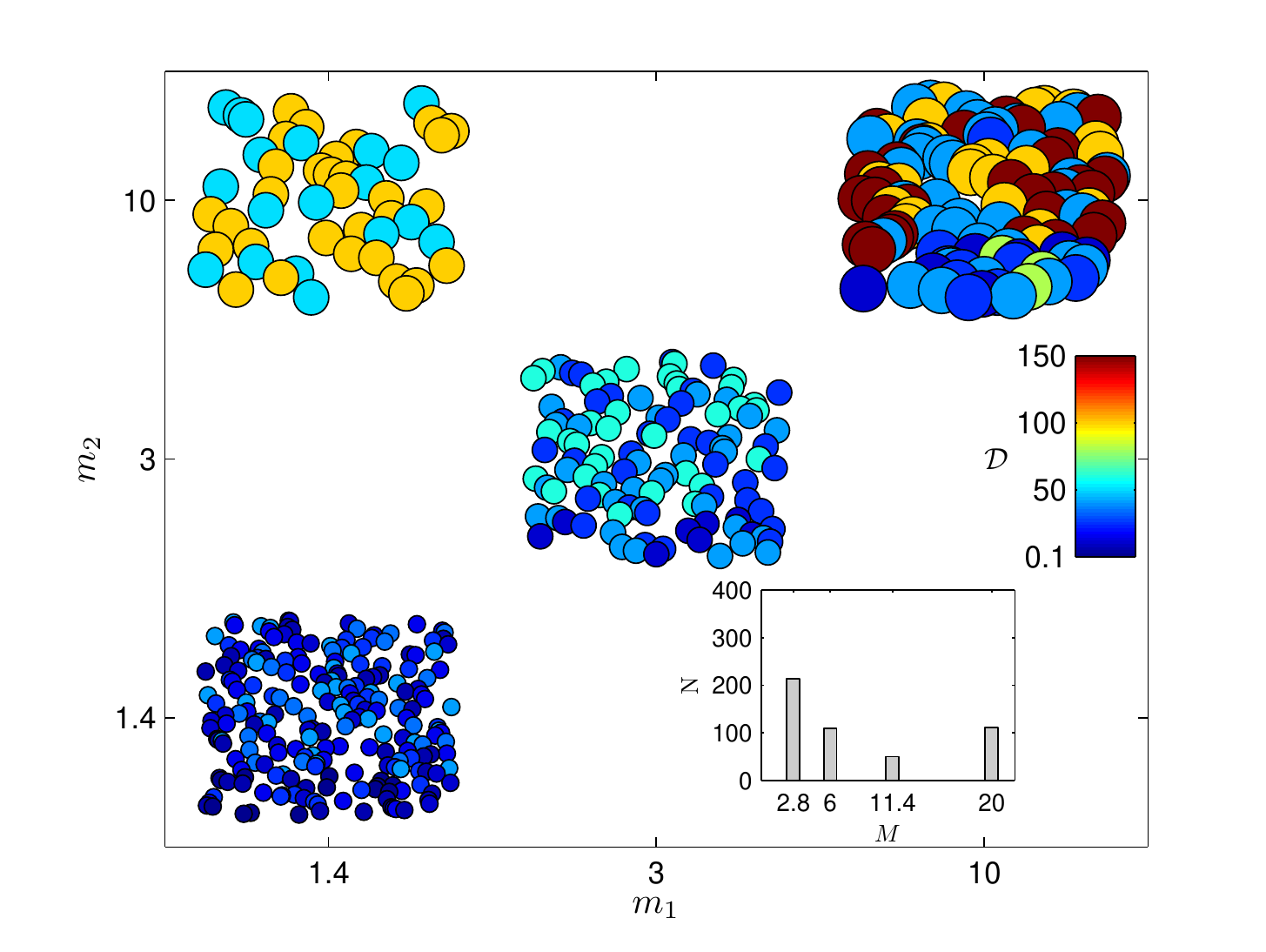}
  	\\
  	(a) & (b)
	\end{tabular}	
	\caption{Distribution of the binary system masses (in units of the sun mass) and distance (in Mpc) that were used to compute the GW that were injected in (a) H1 and (b) L1 during S5. The clusters are scatter plots for binaries with the same masses (e.g. the left lower cluster is for $m_1=1.4M_\odot$ and $m_2=1.4M_\odot$), which are randomly dispersed to avoid overlapping of the mass circles. The size of the circle represents the total mass $M=m_1+m_2$ while the color represents the distance $\mathcal{D}$ in Mpc. The inner histogram in each plot shows the distribution of the total mass $M$ of the binary systems. Note that there is only total mass $M$ of 1.4, 6, 11.4 and 20 $M_\odot$.}
	\label{fig:InjectionsDistributionMasses}
	\end{center}
\end{figure*}

Data blocks from S5 containing a inspiral binary hardware injection were downloaded and analyzed (see https://losc.ligo.org/about/ for a tutorial on how to download LIGO data). Each data block has a duration of 4096 s and a sampling frequency of 4096 Hz. The parameters of the binary system ($m_1$, $m_2$ and $\mathcal{D}$) as well as the coalescence time ($t_c$) instant of the injection are known in advance. Data blocks with missing strain data (caused for instance by temporary malfunctions or irregularities in the detectors), with unavailable injection information (values of masses or distance of the binary system not reported) or labelled with no successful injection (due to uncontrollable situations during the injection process) were discharged and not used in the rest of this work. Table~\ref{Tab:DescriptionDataBlocks} summarizes the total number of data blocks downloaded, discharged and finally used in this study. For all the used data blocks, figure \ref{fig:InjectionsDistributionMasses} shows the distribution masses ($m_1$ and $m_2$) and the distance ($\mathcal{D}$) of the binary system used to compute the GW signal that were injected in the interferometers. The injected GW presented a total mass of 2.8, 6, 11.4 and 20 $M_\odot$, and the distance ranged from 0.1 to 150 Mpc.
\begin{table}[h]
\caption{Total number of 4096s-long data blocks from H1 and L1 that were downloaded, discarded and finally used to assess the detection of GW from binary systems injected in LIGO's S5 run.}
\label{Tab:DescriptionDataBlocks}
\begin{center}
\begin{tabular}{c||cc}
                              	 & \multicolumn{2}{c}{\bf LIGO S5} \\
4096s-long data blocks        	 &  H1    &  L1    \\
\hline  \hline
Downloaded                    	 &  874   &  793   \\
With missing data      	         &  175   &  306   \\
With no successful injection     &   5    &   1    \\
\hline
Total                            &  694   &  486   \\
\end{tabular}
\end{center}
\end{table}

%---------------------------------------
\subsection{Detection Methodology}
\label{subsec:DetectionMethodology}
%---------------------------------------
%
To detect the injected GW from binary systems, a data analysis methodology was proposed where the key element is the matched filter algorithm presented in the previous section. This methodology is based on the following assumptions. First, the noise is non-stationary along the time of the entire S5 run, but it is assumed to be stationary during the time of each data block. This is realistic and allows to compute the noise PSD to perform detection of each injected GW. Second, it is known beforehand that a GW is present on the data bock and the coalescence time also known. This is true for LIGO data with injected GW, however this does not hold in the actual search of GW as there is no way to infer whether a GW is present on the data and even less when the end of the inspiral occurs. Finally, the parameters of the GW source are known. This is also not the case in the real search of GW and a bank of templates is required to explore the parameter space.

\begin{figure*}[h]
	\begin{center}
	\begin{tabular}{c}
  	\includegraphics[width=1.0\textwidth]{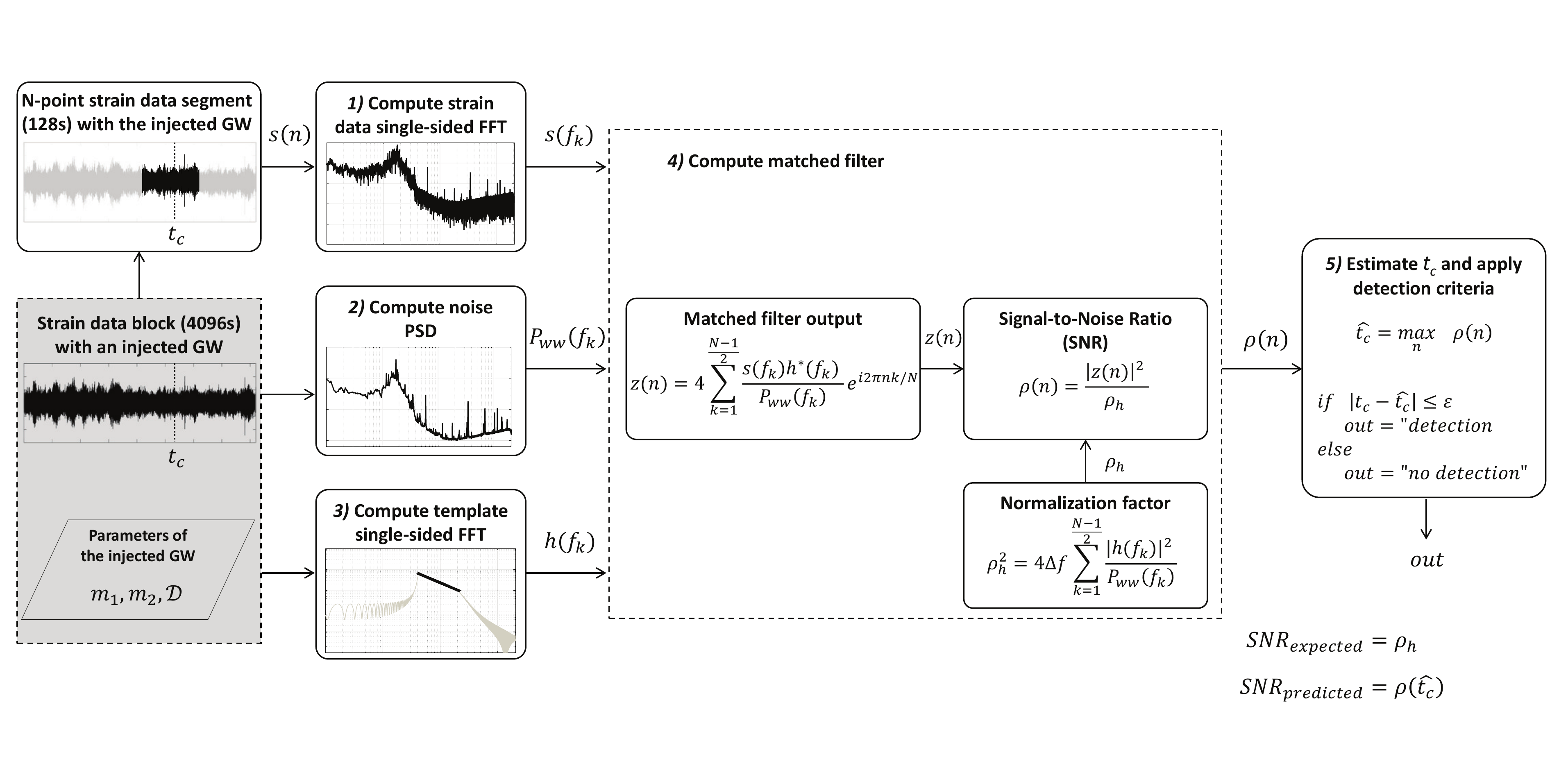} 
	\end{tabular}	
	\caption{Graphical description of the data analysis procedure followed to detect GW from binary systems in the inspiral phase that were injected in the S5 run. This procedure was performed separately for each 4096s-long data block. The assumptions are that the noise is assumed to be stationary, a GW is present, and the parameters of the GW are known.}
    \label{fig:DataAnalysisProcedure}
	\end{center}
\end{figure*}

Figure \ref{fig:DataAnalysisProcedure} illustrates the data analysis methodology carried out to perform the detection of GW from inspiral binary system that were injected in S5 LIGO data. This procedure was individually applied to each 4096s-long data block containing a GW with known masses, effective distance and time of the end of the inspiral. The detection procedure consists of the following five steps.
\begin{enumerate}
\item Extract a 128s-long strain data segment $s(n)$ containing the injected GW. To carry out this, the full 4096s-long data block is divided in consecutive epochs of length 128 s and the epoch containing the injection is selected as the strain data segment. Thus, number of samples in $s(n)$ is $N = 128 \times 4096$. Then, compute the single-sided strain data Fourier transform $s(f_k)$. % with overlap of 64 s and the first epoch containing
\item Compute the noise power spectral density $P_{ww}(f_k)$. The PSD is computed from the full 4096s-long data block using the Welch's averaged modified periodogram method \cite{welch1967use}. Hanning-windowed epochs of length 128 s with overlap of 64 s were used to compute the spectral power in the frequency range between 0 Hz and 2048 Hz at a resolution of 1/128 Hz. % Note that the noise PSD is computed separately for each data block. Thus we are dealing with the non-stationarity of the noise.
\item Compute the single-sided frequency-domain template of the GW strain $\tilde{h}(f_k)$ in the 2-PN order correction using the a priori known masses and effective distance ($m_1$, $m_2$ and $\mathcal{D}$) of the binary system (see Eq. \ref{equ:FrequencyDomainStrain}).
\item Apply the matched filter algorithm, first compute $z(n)$ and then compute the output signal-to-noise ratio $\rho(n)$ using the normalization constant $\rho_{h}$.
\item Estimate the coalescence time $\widehat{t_c}$, which is defined as the time of the maximum output signal-to-noise ratio or $\widehat{t_c} = \underset{n}{\text{maximize}} \,\, \rho(n)$. Then apply the detection criteria to decide if the detection is successful. Here, a detection is achieved or $out="detection"$ if $|t_c - \widehat{t_c}| \leq \varepsilon$ and no detection or $out="no \, detection"$ is obtained otherwise, where $\varepsilon$ is the 10\% of the chirp duration. Finally, compute the expected and the recovered signal-to-noise ratio, $SNR_{expected} = \rho_h$ and $SNR_{recovered} = \rho(\widehat{t_c})$, respectively.
\end{enumerate}

As a result of applying this procedure on each data block, the parameters of the GW ($m_1$, $m_2$ and $\mathcal{D}$) along with the detection results ($out$, $SNR_{expected}$ and $SNR_{recovered}$) are obtained.

%---------------------------------------
\subsection{Detection results}
\label{subsec:DetectionResults}
%---------------------------------------
%
To asses performance in the detection of the known GW, we defined the detection accuracy as the percentage of correct detections or $ACC = Number \, of \, Detections/Number \, of \, Injections$. The detection accuracy was computed across all injected GW and considering all injections of the same total mass. Overall, GW were successfully detected in 88\% and 82\% of injections for H1 and L1, respectively. In addition, detection accuracy varies with respect to the total mass, but no relationship is observed between $M$ and $ACC$.
\begin{table}[h]
\caption{Detection accuracy ($ACC$) results for H1 and L1. The $ACC$ was computed for all injection and considering injections with the same total mass $M$ (in units of solar mass $M_\odot$). % and with the same effective distance $\mathcal{D}$.
}
\label{Tab:DetectionPerformance}
\begin{center}
\begin{tabular}{cp{1cm}c}
\begin{tabular}{c||c|c|c|c||c}
            & \multicolumn{5}{|c}{Total mass (units of $M_\odot$)}         \\
            & \,\,\,$2.8$\,\,\, & \,\,\,$6.0$\,\,\, & \,\,\,$11.4$\,\,\, & \,\,\,$20.0$\,\,\, & \bf \,\,\,All\,\,\, \\
\hline \hline
Injections:                     & 315    & 142    & 77     & 160    & \bf 694   \\
Detections:                     & 264    & 138    & 57     & 150    & \bf 609   \\
\hline
\bf $ACC$ (\%)                  & 84     & 97     & 74     & 94     & \bf 88    \\
\end{tabular}
&

&
\begin{tabular}{c||c|c|c|c|c}
            & \multicolumn{5}{|c}{Total mass (units of $M_\odot$)}         \\
            & \,\,\,$2.8$\,\,\, & \,\,\,$6.0$\,\,\, & \,\,\,$11.4$\,\,\, & \,\,\,$20.0$\,\,\, & \bf \,\,\,All\,\,\, \\
\hline \hline
Injections:                     & 214    & 109    & 50     & 112    & \bf 486   \\
Detections:                     & 162    & 102    & 31     & 105    & \bf 400   \\
\hline
\bf $ACC$ (\%)                  & 76     & 94     & 62     & 94     & \bf 82    \\
\end{tabular}
\\
(a) $ACC$ for H1 & & (b)  $ACC$ for L1 
\end{tabular}
\end{center}
\end{table}

\begin{figure*}[h]
	\begin{center}
	\begin{tabular}{ c c }
  	\includegraphics[width=0.5\textwidth]{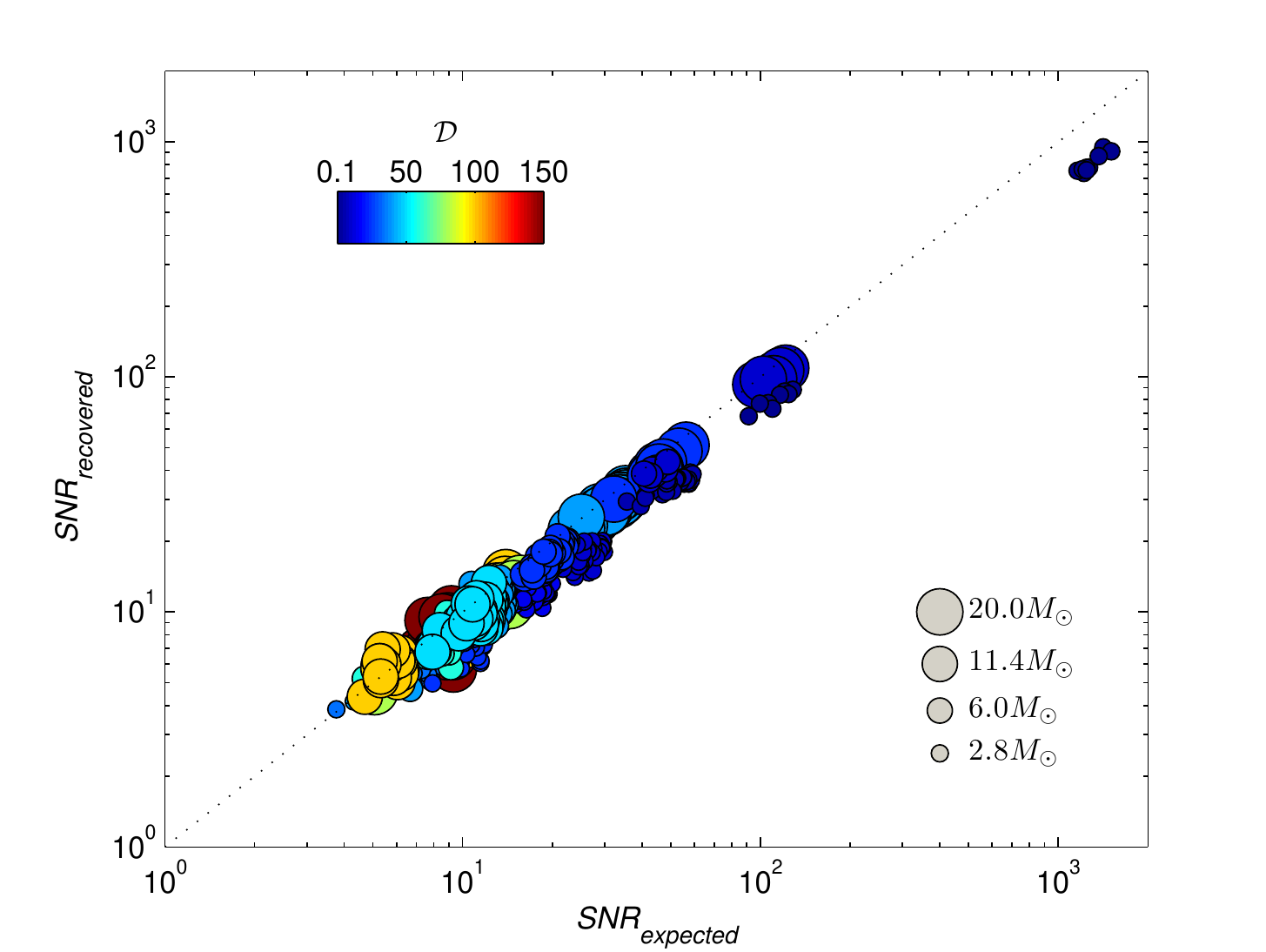} 
  	&
  	\includegraphics[width=0.5\textwidth]{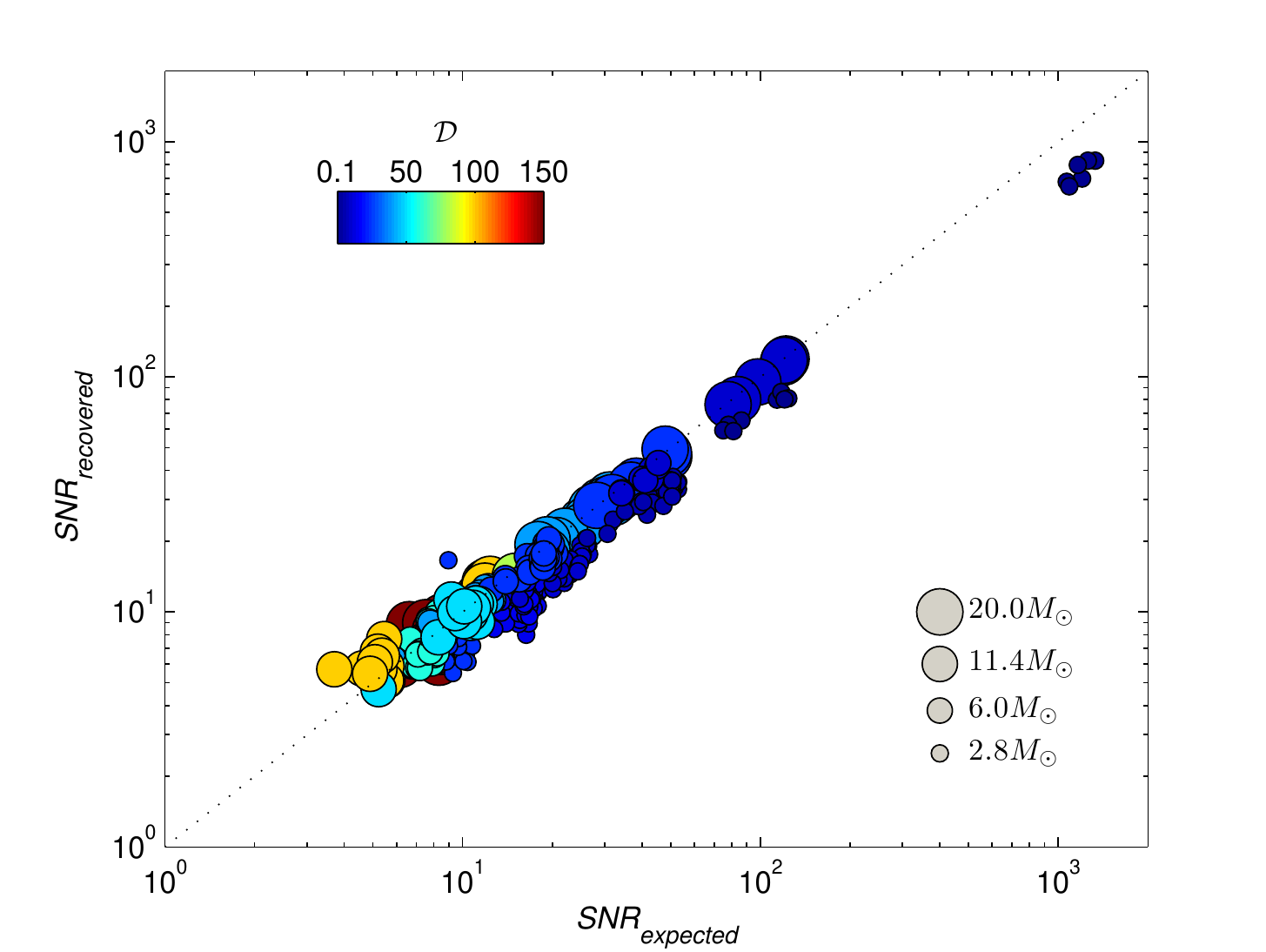}
  	\\
  	\multicolumn{2}{c}{(a) $SNR_{expected}$ versus $SNR_{recovered}$ for all successful detected GW.} 
  	\\
  	\includegraphics[width=0.5\textwidth]{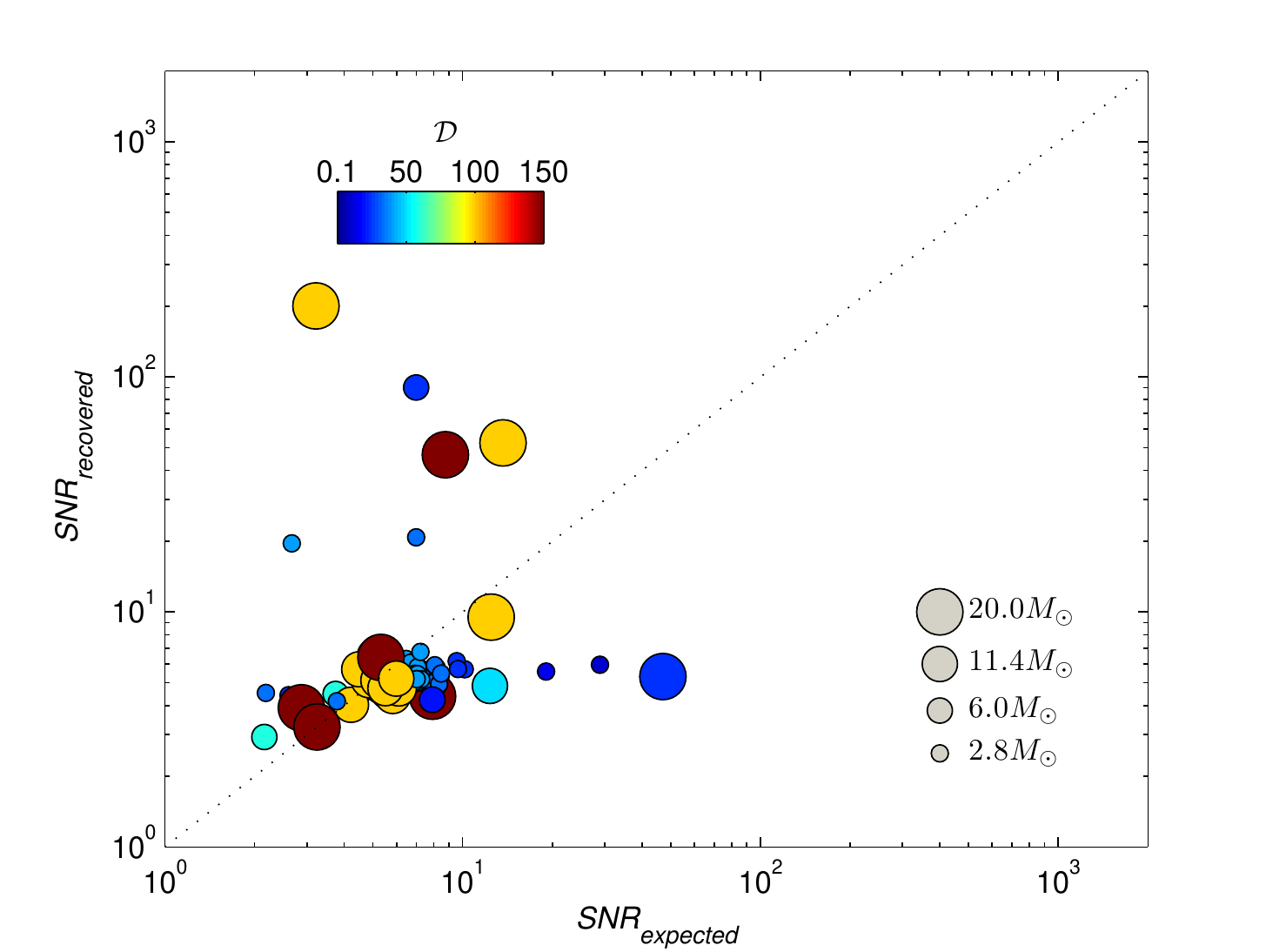} 
  	&
  	\includegraphics[width=0.5\textwidth]{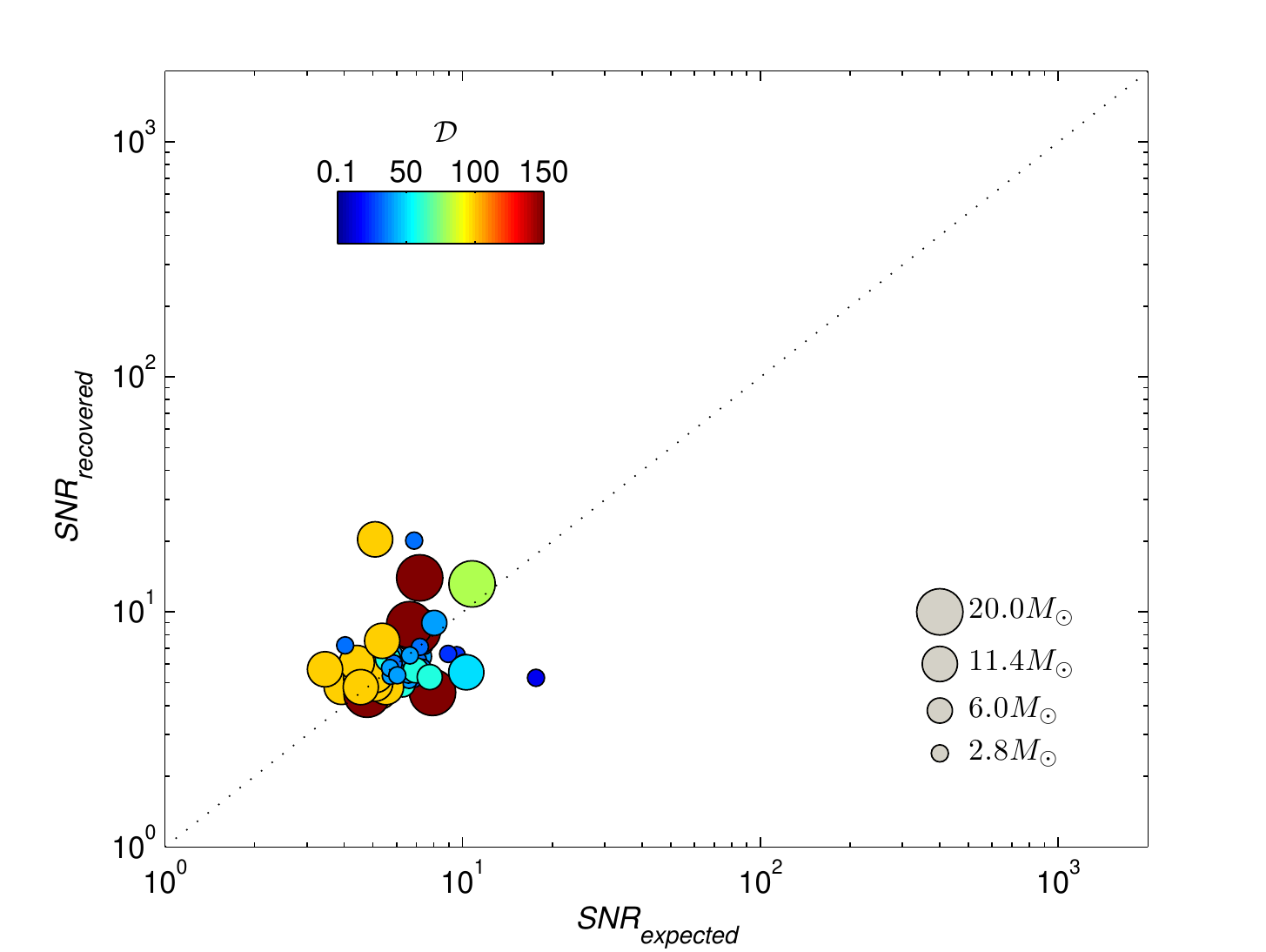}
  	\\
  	\multicolumn{2}{c}{(b) $SNR_{expected}$ versus $SNR_{recovered}$ for all no detected GW.} 
	\end{tabular}	
	\caption{Expected versus recovered signal-to-noise ratio ($SNR_{expected}$ versus $SNR_{recovered}$) in H1 (left panels) and L1 (right panels) for (a) detected and (b) for no detected GW. The size of the circle represents the total mass $M=m_1+m_2$ while the color represents the distance $\mathcal{D}$ in Mpc.}
	\label{fig:ExpectedVersusRecoveredSNR}
	\end{center}
\end{figure*}

The expected versus the recovered signal-to-noise ratio ($SNR_{expected}$ versus $SNR_{recovered}$) is presented in figure \ref{fig:ExpectedVersusRecoveredSNR} separately for detected and for non-detected GW. For successful detected GW (figure \ref{fig:ExpectedVersusRecoveredSNR}a) the recovered SNR is significantly linear correlated to the expected SNR ($R^2=0.9963$ and $p<0.001$ for H1; $R^2=0.9931$ and $p<0.001$ for L1). However, for non-detected GW (figure \ref{fig:ExpectedVersusRecoveredSNR}b) no significant linear correlation was found between the recovered and expected SNR ($R^2=0.0022$ and $p=0.6704$ for H1; $R^2=0.0097$ and $p=0.3734$ for L1). These results show the match between expected and the recovered signal-to-noise ratio only when GW are detected.

\begin{figure*}[h]
	\begin{center}
	\begin{tabular}{ c c }
  	\includegraphics[width=0.5\textwidth]{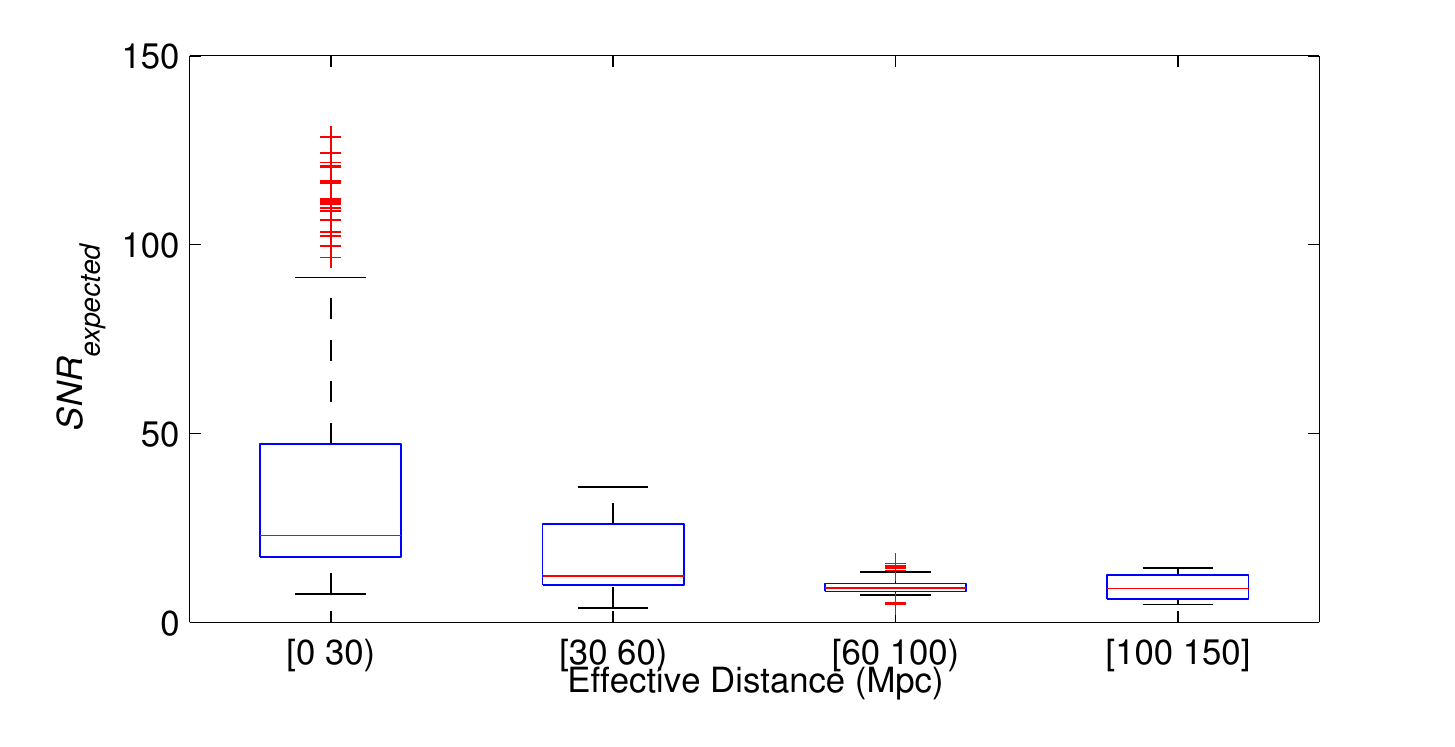} 
  	&
  	\includegraphics[width=0.5\textwidth]{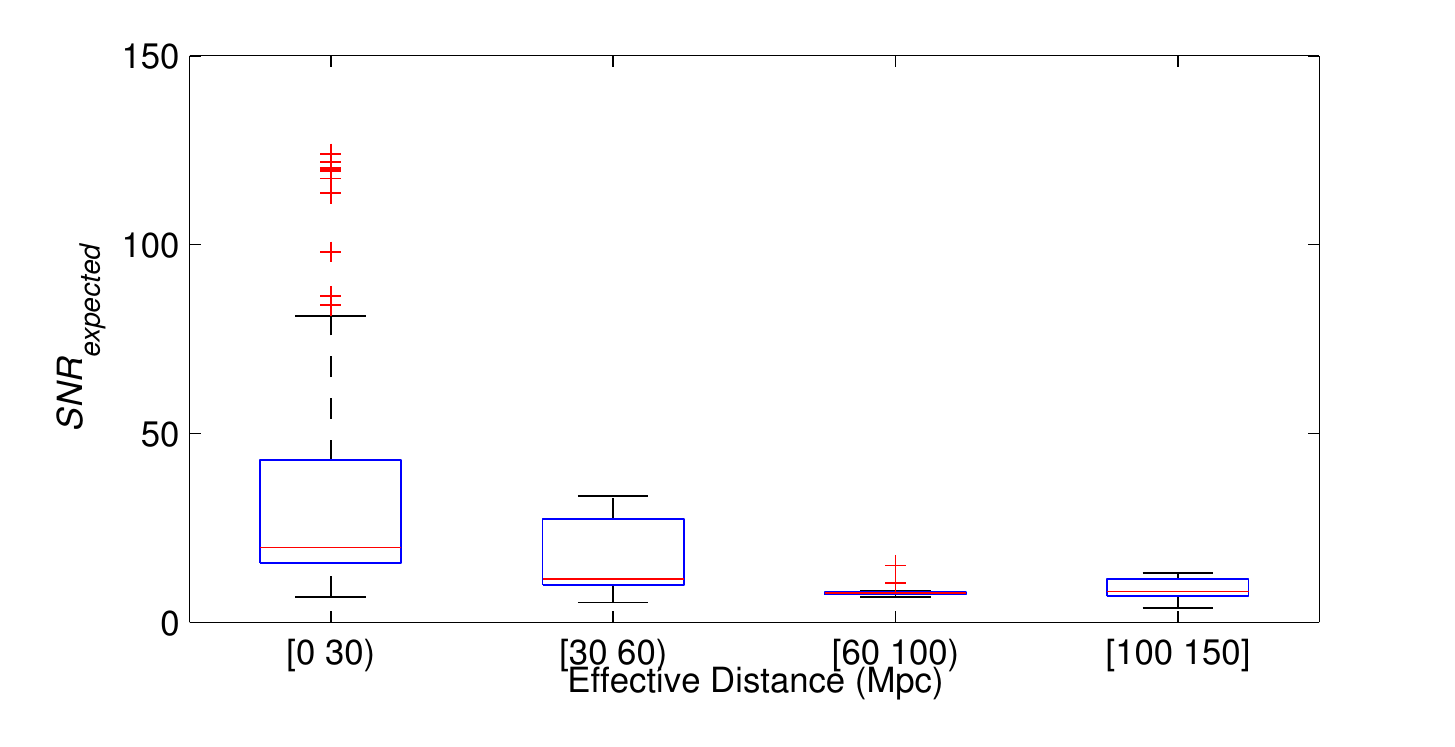}
	\end{tabular}	
	\caption{Distribution of the expected signal-to-noise ratio ($SNR_{expected}$) in H1 and L1 for successful detected GW according to the effective distance of the source.}
	\label{fig:ExpectedSNRVersusDistance}
	\end{center}
\end{figure*}

To examine the relationship between the signal-to-noise ratio and the effective distance of the binary, figure \ref{fig:ExpectedSNRVersusDistance} shows the distribution of the expected signal-to-noise ratio ($SNR_{expected}$) for different ranges of the distance. These results shows that the closer to the source to the detector, the greater the $SNR_{expected}$ that the GW signal would attain in the detector.

%\input{05_Conclusions}
%
%%%%%%%%%%%%%%%%%%%%%%%%%%%%%%%%%%%%%%%
\section{Conclusion}
\label{sec:Conclusion}
%%%%%%%%%%%%%%%%%%%%%%%%%%%%%%%%%%%%%%%
%
Gravitational Waves (GW), the ripples in the fabric of the space-time predicted one hundred years ago in the Einstein's General Theory of Relativity, have been finally detected. These waves are produced by massive astrophysical objects moving at violent accelerations and propagate at the speed of light. The tremendous effort of the LIGO scientific collaboration \cite{0264-9381-23-19-S03,0264-9381-32-7-074001} resulted in the detection of  the events GW150914 and GW151226 which were emitted by stellar-mass binary systems composed of two merging black holes \cite{Abbott:2016,Abbott:2016b}. These discoveries have gave new life to the General Relativity \cite{Misner73} opening a new era for astronomy, a new alternative to study and to understand the universe, and the expectation of novel technological developments \cite{sR00}.

Due to the importance of this discovery there is a need to clearly understand both the theoretical and practical foundations of GW. For this reason, this work first presents an introduction to the fundamental theory of GW focusing on the mathematical model of GW generated by binary systems in the inspiral phase. In this case, the post-Newtonian limit allows to compute an analytical solution yielding to a GW signal that exhibit a chirp-type waveform. Then, it is briefly described the scientific and technological efforts developed to detect GW using LIGO. Secondly, the work proposes a comprehensive data analysis methodology developed to detect GW emitted by inspiral binary systems which are embedded in the recorded data from LIGO. The methodology is based on the matched filter algorithm, which is the standard approach to search GW signals whose waveform is known in advance and basically correlates in the frequency domain the observed LIGO data with the waveform of the expected GW over the detector's sensitive band.

The proposed method was validated with freely available LIGO data from the fifth science run (S5) during which the detectors operated at the initial design sensitivity. Although no GW were detected, the recorded data contain simulated GW signals from binary systems (created based on the 2-order PN approximation) which were injected during the recording process. Several hundred of data blocks from both LIGO detectors containing an injected GW were used to search and to assess detection of GW waveforms. The detection accuracy results showed that GW were detected in roughly 85\% of the cases across the two LIGO detectors. This result shows that the majority of the GW were successfully recovered. The detection accuracy was also computed separately for GW generated by sources with the same total mass, however, the results showed no relationship between them. This is because despite the total mass is the same, the distance at which the GW source is located varies in a broad range. Finally, the analysis of the expected versus the recovered signal-to-noise ratio showed significant linear correlation for successful detected GW, while no significant linear correlation was found for no detected GW. These results revealed the match between expected and the recovered signal-to-noise ratio only when GW are detected.

It is important to point out that the data analysis methodology presented herein assumes that the detector noise is stationary during the time period where the search of GW is performed, moreover, it is required to known beforehand that a GW is present along with its parameters (masses of the binary companion and the distance of its location). Therefore, these assumptions hold for LIGO data with injected or confirmed events of GW. Notice that in the actual search of GW it is unknown if there is a GW and thus its parameters are also unknown. In this case, the search is carried out over a template bank containing many GW signals that cover the parameters space, and detection involves applying the matched filter and statistical tests to discharge spurious detections.

The next step in this research is to apply this method to search: $(i)$ GW signals injected during sixth science run (S6), where the GW were generated using the 3.5-PN order and the detector operated and the enhanced sensitivity; and $(ii)$ the events GW150914 and GW151226, where the GW waveform models can be extracted from several available catalogues and the detector operated and the final advanced sensitivity. In addition we will explore our own detection using GW waveform models that include spin of the objects in the binary. This constitutes an enhanced search and provides the opportunity of discovery GW that have not been detected before.

%%%%%%%%%%%%%%%%%%%%%%%%
%%% ACKNOWLEDGMENT   %%%
%%%%%%%%%%%%%%%%%%%%%%%%
%
%\acknowledgments
%This research was supported by CONACyT-AEM Grants No. 248411 and 262847. CM want to thanks at the Universidad de Guadalajara for academic and financial support.

%%%%%%%%%%%%%%%%%%%%%%%%
%%% ACKNOWLEDGMENTS  %%%
%%%%%%%%%%%%%%%%%%%%%%%%
%
\section*{Acknowledgements}
This research has made use of data, software and/or web tools obtained from the LIGO Open Science Center (https://losc.ligo.org), a service of LIGO Laboratory and the LIGO Scientific Collaboration. LIGO is funded by the U.S. National Science Foundation. The authors would like to thank the support of the CONACyT grant No. 271904 and the CONACyT-AEM grants No. 248411 and 262847. CM want to thank PROSNI-UDG 2016.

%%%%%%%%%%%%%%%%%%%%%%%%
%%%    REFERENCES    %%%
%%%%%%%%%%%%%%%%%%%%%%%%
%
%\bibliographystyle{unsrt}
\bibliography{References}

%%%%%%%%%%%%%%%%%%%%%%%%
%%%       END        %%%
%%%%%%%%%%%%%%%%%%%%%%%%
%
\end{document}